\documentclass[9pt,twocolumn,twoside]{pnas-new}

\templatetype{pnasresearcharticle} 

\title{A random critical point separates brittle and ductile yielding transitions in amorphous materials}

\author[a]{Misaki Ozawa}

\affil[a]{Laboratoire Charles Coulomb (L2C), Universit\'e de Montpellier, CNRS, Montpellier, France}

\author[a]{Ludovic Berthier}

\author[b,c]{Giulio Biroli}

\affil[b]{Institut de physique th\'eorique, Universit\'e Paris Saclay, CEA, CNRS, F-91191 Gif-sur-Yvette, France }

\affil[c]{Laboratoire de Physique Statistique, \'Ecole Normale Sup\'erieure, CNRS, PSL Research University, Sorbonne Universit\'e, 75005 Paris, France}

\author[d]{Alberto Rosso}

\author[e]{Gilles Tarjus}

\affil[d]{LPTMS, CNRS, Univ. Paris-Sud, Universit\'e Paris-Saclay, 91405 Orsay, France}

\affil[e]{Laboratoire de Physique Th\'eorique de la Mati\`ere Condens\'ee, CNRS UMR 7600,
UPMC-Sorbonne Universit\'e, 4, place Jussieu, 75252 Paris Cedex 05, France}

\leadauthor{Misaki Ozawa} 

\significancestatement{Understanding how amorphous solids yield in response to external deformations is crucial both for practical applications and for theoretical reasons. Here we show that despite large differences in the materials microscopic structures, a degree of universality emerges as there are only two ways in which amorphous solids respond to a deformation: One, typical of well annealed materials, is characterized by an abrupt failure with a macroscopic stress drop and the sudden emergence of sharp shear bands; the other, typical of poorly annealed materials, is smooth and only shows a plastic regime. By varying the preparation protocol, one can change the response of a given material from one to the other, and this change is controlled by a random critical point.}

\authorcontributions{All authors contributed equally to this work.}

\begin{abstract}
We combine an analytically solvable mean-field elasto-plastic model with molecular dynamics simulations of a generic glass-former to demonstrate that, depending on their preparation protocol, amorphous materials can yield in two qualitatively distinct ways. We show that well-annealed systems yield in a discontinuous brittle way, as metallic and molecular glasses do. Yielding corresponds in this case to a  first-order nonequilibrium  phase transition. As the degree of annealing decreases, the  first-order character becomes weaker and the transition terminates in a second-order critical point in the universality class of an Ising model in a random field. For even more poorly annealed systems, yielding becomes a smooth crossover, representative of the ductile rheological behavior generically observed in  foams, emulsions, and colloidal glasses. Our results show that the variety of yielding behaviors found in amorphous materials does not necessarily result from the diversity of particle interactions or microscopic dynamics, but is instead unified by carefully considering the role of the initial stability of the system.
\end{abstract}

\dates{\today}
\doi{\url{www.pnas.org/cgi/doi/10.1073/pnas.XXXXXXXXXX}}

\begin{document}

\verticaladjustment{-2pt}
\maketitle
\thispagestyle{firststyle}
\ifthenelse{\boolean{shortarticle}}{\ifthenelse{\boolean{singlecolumn}}{\abscontentformatted}{\abscontent}}{}




\section{Introduction}

In amorphous solids, yielding generically signals a macroscopic change of the global mechanical response from an apparent elastic-solid regime at small deformation to a plastic-flow regime at large deformation~\cite{bonn2017yield,nicolas2017deformation,schuh2007mechanical}. Yielding can occur smoothly, as when one spreads cream cheese on a bagel, or can be abrupt and catastrophic, as when a smartphone screen breaks. Yielding is observed in soft glasses such as emulsions, colloidal suspensions, and foams~\cite{bonn2017yield}, but also in molecular and metallic glasses~\cite{schuh2007mechanical}. It represents a central problem in statistical physics~\cite{nicolas2017deformation} (can yielding be described as a nonequilibrium phase transition?), soft condensed matter~\cite{bonn2017yield} (how do soft glasses flow?), and materials science~\cite{schuh2007mechanical} (can one predict material failure?). Understanding the fate of an amorphous material that is mechanically driven very slowly from an initial disordered configuration represents the core challenge, and its solution would directly allow one to understand steady-state flows~\cite{bonn2017yield}, oscillatory deformations~\cite{knowlton2014microscopic}, shear-band formation~\cite{greer2013shear} and, perhaps most importantly, the catastrophic failure of amorphous solids~\cite{schuh2007mechanical}. 

Failure and flow of a disordered solid is such a ubiquitous phenomenon in nature and technological applications that it has stimulated an intensive search for universal explanations~\cite{reche2008,cubuk2017structure,uhl2015universal}. One such explanation is based on elasto-plastic models~\cite{picard04elastic,picard2005slow,nicolas2017deformation,cao2017soft} and their analogy with the  depinning of a manifold in a random environment~\cite{wyart-rosso,baret2002extremal}; it has recently allowed a clarification of the critical nature of the steady-flow regime observed at very large deformation. In this stationary regime, the stress undergoes a succession of elastic charges interrupted by sudden plastic drops. Microscopically, plasticity corresponds to localized particle rearrangements, called shear transformations \cite{argon1979plastic,falk1998dynamics}, which release the accumulated stress and induce long-range reorganization triggering system-spanning avalanches. Universality emerges because the stress drops display scale-free statistics, similar to the Gutenberg-Richter law for earthquakes \cite{wyart-rosso,baret2002extremal,dahmen1,salerno2013effect,maloney2006amorphous,nicolas2014spatiotemporal}. 

The above studies are focused on "ductile" rheological responses observed in most soft glassy materials (such as cream cheese), which do reach a steady state. Yet, many amorphous solids (such as smartphone screens) are instead "brittle" and fail macroscopically after a finite deformation. For both ductile and brittle materials, the nature of the yielding transition between an elastic-like and a plastic behavior is an actively  studied and vigorously debated question. Different views have been proposed. Yielding has been first described as a spinodal (i.e., a limit of stability) in \cite{wolynes2012} on the basis of random first-order transition theory. Later, in agreement with infinite dimensional computations~\cite{rainone,urbaniz,urbanib}, yielding has been interpreted as a discontinuous transition~\cite{jaiswal2016mechanical}, and then, later on, associated to a critical spinodal~\cite{procacciaspi},   independently of the initial preparation. In addition to the specific characterization of the yielding transition, a crucial open question is why, in spite of their strong structural similarities, are some materials brittle and others ductile? 

Here we show that yielding and brittleness are two facets of the same problem, which we describe at once\footnote{In this paper we use the term “brittle” to characterize a discontinuous yielding. Although this phenomenon is not accompanied by the formation of regions of vacuum, as it happens in the fracture of brittle materials, the macroscopic avalanche taking place at the discontinuous yielding transition does resemble a crack induced by a brittle fracture. In this sense the behavior at   discontinuous yielding is brittle-like.}. We provide a theoretical and numerical analysis of the transient response to an athermal shear deformation starting from the disordered solid. Our main finding is that there are two different universal behaviors, depending on the degree of annealing of the initial configuration but not on the detailed nature of the material {\it per se}. We show that the evolution of the stress caused by quasi-static deformations qualitatively changes from a sharp discontinuous transition when the material is initially well annealed, and therefore very stable, to a smooth crossover as the degree of annealing decreases, and the material is initially less stable. These two regimes are separated by a {\it random critical point}, by which we denote a critical point controlled by the presence of quenched disorder. It is reached for a critical value of the degree of annealing. Our analysis suggests that this criticality is related to the universality class of an Ising model in a random field (generically denoted by RFIM~\cite{nattermann1998theory}).  In this picture, the yielding of ductile materials, which are viewed as rather poorly annealed systems, does not correspond to a genuine phase transition. 

The starting point of our work is the idea, inspired by random first-order transition theory~\cite{wolynes2012} and mean-field calculations~\cite{rainone,urbanib,jaiswal2016mechanical}, that yielding corresponds to a spinodal instability, but we additionally take into account several important features that can change the picture drastically: (i) the presence of quenched disorder, physically corresponding to the intrinsic structural heterogeneity present in amorphous materials, (ii) the possible disappearance of the spinodal that can be replaced by a smooth crossover, and (iii) finite-dimensional fluctuations, which are generically expected to destroy the criticality associated to a mean-field spinodal instability. 
In the following we first support our claims by studying an analytically solvable mean-field elasto-plastic model that we devise to capture the brittle-to-ductile transition through a random critical point. We then use molecular dynamics simulations of a glass-former prepared over an unprecedented range of initial stability, building on very recent computational developments~\cite{ninarello2017models}. The simulations fully confirm the theoretical scenario and provide direct evidence for a random critical point controlling the brittleness of amorphous solids.

\section{Mean-Field Theory} 

To substantiate our proposal we develop a simple analysis, which is inspired by the description of sheared materials in terms of elasto-plastic models~\cite{nicolas2017deformation}. This widespread mesoscopic approach successfully reproduces the key phenomenology of deformation and flow in amorphous materials. Our main focus is on the role of the initial preparation, which has received much less attention  (see however \cite{vandembroucq2011mechanical,lin2015criticality,vasoya2016notch,liu2018mean}).  

In this approach the system is decomposed in mesoscopic blocks $i=1,\cdots,N_{\rm b}$, in which elastic behavior is interrupted by sudden shear transformations.
At each block is assigned a local stress, $\sigma_i$, drawn from an initial distribution $P_{\text{ini}}(\sigma)$ which encodes the degree of annealing.  In the absence of plastic events the response is purely elastic, and a small deformation increment, $\delta \gamma$, loads all the blocks as $\sigma_i \to \sigma_i+ 2 \mu_2 \delta \gamma$, with $\mu_2$ the shear modulus. However, when the local stress becomes larger than a threshold value $\sigma_i^{\rm th}$ (that for simplicity we consider uniform $\sigma_i^{\rm th}=\sigma^{\rm th}$), the block yields and the local stress  drops by a random quantity $x\geq 0$ sampled from a given distribution $g(x)$. After the drop, the stress is redistributed to the other sites as $\sigma_j \to \sigma_j +{\mathcal G}_{ij} x$. The elastic kernel ${\mathcal G}_{ij}$ is generally taken of the Eshelby form which corresponds to the far-field solution of elasticity (it decays as $1/|i-j|^d$, where $d$ is the spatial dimension, but changes sign and displays a quadrupolar symmetry)~\cite{eshelby1957determination}. There is no straightforward and generally accepted way to handle the nonlocal Eshelby interaction kernel at a mean-field level~\cite{hebraud1998mode,caroli-lemaitreII,lin-wyart}. Here we consider a mean-field approximation that consists in replacing this nonlocal interaction by a fully connected kernel ${\mathcal G}_{ij} = \frac{\mu_2}{N_{\rm b}(\mu_1+\mu_2)}$, with $\mu_1>0$. This description overlooks the effect of the anisotropic and nonpositive form of the Eshelby interaction kernel. Nonetheless, we expect that it provides a correct qualitative description of the yielding transition itself. (A similar behavior is indeed found by analyzing more involved mean-field models~\cite{wyart18}.) Below we discuss its limitations and how to go beyond them. Note that this model has also a natural interpretation as a mean-field model of depinning\footnote{A narrower initial distribution, different from the stationary one, corresponds to aging in the quenched disordered~\cite{jagla07shear,vandembroucq2011mechanical}, i.e., to a stronger pinning at initial times.} as well as earthquake statistics~\cite{fisher1997statistics,jagla2014viscoelastic}.

The key quantity in this approach is the distribution $P_\gamma(x)$ of the distances $x_i=\sigma^{\rm th}-\sigma_i$ from the threshold stress. In the following we study its macroscopic evolution with strain $\gamma$. As detailed in the SI, it is governed by the equation
\begin{equation}\label{eqpx}
\frac{\partial P_\gamma(x)}{\partial \gamma}=  \frac{2 \mu_2}{1-x_c P_\gamma(0)} \left [ \frac{\partial P_\gamma(x)}{\partial x} + P_\gamma(0) g(x) \right ] ,
\end{equation}   
where $x_c=(\mu_2/[\mu_1+\mu_2])\bar x$ and $\bar x =\int_0^{\infty} d x x g(x)$ represent material-dependent parameters (here, we have $x_c<\bar x<1$ as we set $\sigma^{\rm th}= 1$ as stress unit).  The degree of annealing of the material is fully encoded in the initial distribution $P_{\gamma=0}(x)$, which contains the same information as $P_{\rm ini}(\sigma)$.

\begin{figure}[t]
\includegraphics[width=0.95\linewidth]{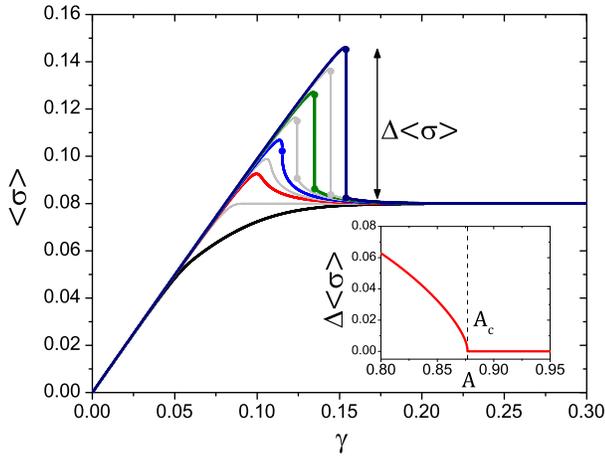}
\caption{The different yielding regimes in the mean-field elasto-plastic model. Stress $\langle \sigma \rangle$ versus strain $\gamma$ for increasing degree of annealing (decreasing A values) from bottom to top, using $x_c=0.9, \bar x =0.92$. The monotonic flow curve (black) transforms into a smooth stress overshoot (red), and above a critical point with infinite slope (blue) becomes a discontinuous transition (green) of increasing amplitude (dark blue). 
Inset: Stress discontinuity $\Delta\langle \sigma \rangle$  versus the degree of annealing (here, the initial distribution $P_{\gamma=0}(x)$ is parametrized by a single  parameter  $A$ which plays the same role as the preparation temperature in the simulations).
}
\label{fig_MF}
\end{figure}

The properties of the macroscopic stress-strain curves can be obtained through Eq.~(\ref{eqpx}) and the relation $\langle \sigma \rangle= 1- \langle x \rangle$, which is derived by taking the average of the equation defining $x_i$. Our results, which hold for a {\it generic} $g(x)$ (see the SI), are shown in Fig.~\ref{fig_MF} for the explicit case $g(x)=\exp(-x/\bar x)/\bar x$ and  $P_{\gamma=0}(x)=\left(e^{-x/A}-e^{-x/(1-A)}\right)/(2A-1)$, $1/2<A<1$. With this choice, $A$ is the unique parameter controlling the degree of annealing, smaller values of $A$ corresponding to better annealed samples.

For a poor annealing, the stress-strain curve is monotonically increasing and yielding is a mere crossover. As one increases the degree of annealing, a stress overshoot first appears, but yielding remains a crossover, still not a {\it bona fide} phase transition. For the best annealing, the overshoot is followed by a spinodal and a sharp discontinous jump of the average stress. Mathematically, this occurs when, increasing $\gamma$, $P_\gamma(0)$ reaches $x_c^{-1}$, thus inducing a singular behavior of $P_\gamma(x)$ via Eq.~(\ref{eqpx}). In this case, yielding takes place as a nonequilibrium first-order transition. Crucially, a critical point $A_c$ separates the first-order regime from the smooth one. From Fig.~\ref{fig_MF} it is clear that an appropriate order parameter distinguishing the two regimes of yielding is the macroscopic stress drop $\Delta \langle \sigma \rangle$. As shown in the inset of Fig.~\ref{fig_MF}, $\Delta \langle \sigma \rangle$ vanishes at large $A$, but it grows continuously by decreasing $A$ below $A_c$. This critical point is therefore reached not only for a specific value of the strain and stress, but also by tuning the degree of annealing of the material. The stress overshoot, frequently observed in colloidal materials~\cite{koumakis2012yielding,amann2013overshoots}, yield stress fluids \cite{divoux2011stress} and computer simulations~\cite{rodney2011modeling}, is simply a vestige of this critical point at larger disorder strength. 

When a spinodal, followed by a discontinuity, is present, the stress displays a square-root singularity as the yield strain $\gamma_Y$ is approached from below, and the distribution of the avalanche size $S$ becomes for large $S$:
\begin{eqnarray} \langle \sigma \rangle -\sigma_{\rm sp} &\propto& (\gamma_Y-\gamma)^{1/2}, \\ \mathcal P(S) &\sim& S^{-3/2} e^{-C (\gamma_Y-\gamma)S},
\end{eqnarray}
where $C>0$ is a constant and $\gamma \to \gamma_Y^-$. The discontinuous stress drop decreases as annealing becomes poorer, and it eventually vanishes with a square-root singularity at the critical point ($\gamma_{Y_c},\sigma_c$), where  
a different behavior emerges:  
\begin{eqnarray}
\langle \sigma \rangle - \sigma_{c} &\propto& {\rm sgn}(\gamma-\gamma_{Y_c})|\gamma-\gamma_{Y_c}|^{1/3},\\ \mathcal P(S) &\sim& S^{-3/2} e^{-C'|\gamma-\gamma_{Y_c}|^{4/3}S},
\end{eqnarray}
with $C'>0$ and $\gamma \to \gamma_{Y_c}^\pm$. All these scaling behaviors coincide with those found for the RFIM\footnote{More precisely,  in the present context, one should consider the out-of-equilibrium behavior of the RFIM when quasi-statically driven at zero temperature by a change of the applied magnetic field~\cite{sethna}. At the mean-field level, this critical behavior is the same as that of the RFIM in equilibrium.} within the mean-field theory~\cite{dahmen-sethna}. 

The presence of an {\it annealing-controlled  random critical point} is the main finding of our mean-field approach. We stress that its presence, as well as that of the different regimes of yielding, does not require the introduction of any additional physical mechanism, such as dynamical weakening~\cite{fisher1997statistics, dahmen2009micromechanical, jagla07shear,jagla2014viscoelastic,papanikolaou2012quasi,patrick17}. It only depends on the initial preparation of the amorphous material prior to shearing, in combination with the basic rules of elasto-plastic models. In finite dimensions, the above scaling behaviors will be modified. Whereas a spinodal instability can still be present in athermal conditions, it will likely not be associated to any critical behavior~\cite{nandi}. On the other hand, the random critical point should always be in the universality class of the athermally driven RFIM, but this class is presumably distinct from that of the conventional model with only short-ranged ferromagnetic interactions.

This mean-field description is not meant to reproduce all aspects of the deformation-and-flow phenomenology. In particular, it does not allow criticality of the sheared system along the elastic and plastic branches~\cite{muller2015marginal,lin2015criticality}, nor can it describe spatial flow inhomogeneities, such as shear bands. Nonetheless, as we now show by computer simulations, the model correctly captures the preparation-dependence of the yielding transition, the central question addressed by our work.

\section{Atomistic model and numerical procedures}

We have numerically studied the yielding transition in a three-dimensional atomistic glass model for different degrees of annealing, with our mean-field predictions as a guideline. We have used a size-polydisperse model with a soft repulsive potential~\cite{ninarello2017models}. Glass samples have been prepared by first equilibrating liquid configurations at a finite temperature, $T_{\rm ini}$, and then performing a rapid quench to $T=0$, temperature at which the samples are subsequently deformed. The preparation temperature $T_{\rm ini}$ then uniquely controls the glass stability, and we consider a wide range of preparation temperatures, $T_{\rm ini}=0.062 - 0.200$. In order to obtain well annealed systems, we have used the swap Monte Carlo (SWAP) algorithm that allows equilibration at extremely low temperatures~\cite{ninarello2017models}.
The considered range of $T_{\rm ini}$ describes very poorly annealed glasses ($T_{\rm ini} \approx 0.2$, corresponding to wet foam experiments), ordinary computer glasses ($T_{\rm ini} \approx 0.12$, corresponding to colloidal experiments), well annealed glasses ($T_{\rm ini} \approx 0.085 - 0.075$, corresponding to metallic-glass experiments), as well as  ultrastable glasses ($T_{\rm ini} \approx 0.062$, see \cite{fullerton2017density}). No previous numerical work has ever accessed such a large range of glass stability. 

We have performed strain-controlled athermal quasi-static shear (AQS) deformation using Lees-Edwards boundary conditions~\cite{maloney2006amorphous}. Note that during the AQS deformation, the system is always located in a potential energy minimum, such that inertia and thermal fluctuations play no role. This method is considered as the zero-strain rate limit, $\dot{\gamma} \to 0$, which bypasses the timescale gap between simulation and experiments~\cite{rodney2011modeling}. Thus, our simulation setting (SWAP / AQS) fully overcomes the timescale gap in both glass preparation and mechanical deformation between simulations and experiments. In order to study the finite-size effect, we have varied the number of particles $N$ over a considerable range, $N=1500-96000$. More details are given in the SI.
 
\section{The two regimes of yielding} 
\label{sec:two_yielding}

\begin{figure}[t]
\centering
\includegraphics[width=1.\linewidth]{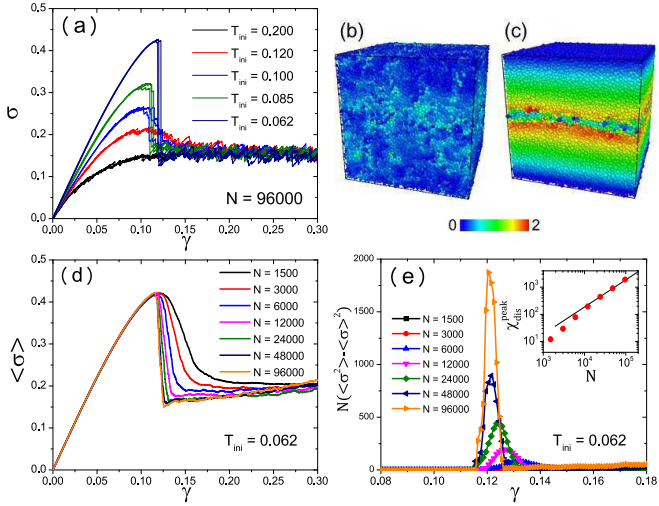}
\caption{(a): The yielding regimes in the simulation of a sheared glass for different degrees of annealing.
Stress $\sigma$ as a function of the strain $\gamma$ for several preparation temperatures $T_{\rm ini}$. For each $T_{\rm ini}$, three independent samples are shown.
(b, c): Snapshots of non-affine displacements between $\gamma=0$ and yielding at $\gamma=0.13$ for $T_{\rm ini}=0.120$ (b) and at $\gamma=0.119$ for $T_{\rm ini}=0.062$ (c). 
(d, e): Evidence of a first-order yielding transition for well annealed glasses. System-size dependence of the averaged stress-strain curve for $T_{\rm ini}=0.062$, showing a sharper stress drop for larger $N$ (d). The associated susceptibility, $\chi_{\rm dis}=N(\langle \sigma^2 \rangle - \langle \sigma \rangle^2)$ becomes sharper as $N$ increases (e). Inset: The divergence of the maximum of $\chi_{\rm dis}$ is propotional to $N$ shown with the straight line.}
\label{fig:simulation_regimes}
\end{figure}

In Fig.~\ref{fig:simulation_regimes}(a), we show the evolution of typical stress-strain curves for large individual samples with $N=96000$. For a high $T_{\rm ini}$, the usual jerky succession of stress drops is found with no overshoot for $T_{\rm ini}=0.200$ (akin to wet foam experiments~\cite{lauridsen2002shear,cantat2006stokes}), and with a stress overshoot for $T_{\rm ini}=0.120$ (akin to colloids experiments~\cite{koumakis2012yielding,amann2013overshoots}). 
Note that previous simulation studies about effects of annealing on yielding behavior would be restricted by this regime~\cite{utz2000atomistic,rodney2011modeling,fan2017effects}. 
Strikingly, the stress overshoot transforms into a sharp stress discontinuity for $T_{\rm ini} \lesssim 0.1$ near the yield strain $\gamma_Y \approx 0.12$. This large stress drop is distinct from the smaller stress drops observed at other strain values. This is confirmed in Fig.~\ref{fig:simulation_regimes}(d), where we plot for increasing values of $N$ the averaged stress $\langle \sigma \rangle$ (obtained by averaging over many  independent samples) for $T_{\rm ini}=0.062$. The stress discontinuity at yielding is the only one surviving after the average and it becomes sharper and better resolved as $N$ increases. These data strongly suggest that, in the thermodynamic limit, the averaged stress-strain curve has a sharp discontinuity at yielding and is smooth everywhere else. This discontinuity is a signature of a non-equilibrium first-order transition, as confirmed by the growth of the associated susceptibilities, the so-called "connected" susceptibility $\chi_{\rm con} = - \frac{d \langle \sigma \rangle}{d \gamma}$ and "disconnected" susceptibility
$\chi_{\rm dis}=N(\langle \sigma^2 \rangle - \langle \sigma \rangle^2)$. The peaks of the susceptibilities become sharper and their amplitude, $\chi_{\rm con}^{\rm peak}$ and $\chi_{\rm dis}^{\rm peak}$, increases with $N$ with exponents expected for a first-order transition in the presence of quenched disorder, as discussed below. This is illustrated for $\chi_{\rm dis}$ in Fig.~\ref{fig:simulation_regimes}(e), and we find that  $\chi_{\rm dis}^{\rm peak} \sim N$ (inset) and $\chi_{\rm con}^{\rm peak} \sim \sqrt{N}$ (see the SI) at large $N$.

The similarity between the mean-field theory in Fig.~\ref{fig_MF} and the data in Fig.~\ref{fig:simulation_regimes}(a) is patent. In agreement with the mean-field theory, we indeed find two distinct types of yielding; a discontinuous one for well annealed glasses, which is associated with a first-order transition that becomes weaker as the degree of annealing decreases, and a continuous one, corresponding to a smooth crossover, for poorly annealed materials. As discussed in the next section, we also find a critical point at $T_{{\rm ini},c} \approx 0.095$ that marks the limit  between the two regimes.

In addition, the simulations give direct real-space insight into the nature of yielding. We illustrate the prominent difference between the two yielding regimes in the snapshots of non-affine displacements measured at yielding in Figs.~\ref{fig:simulation_regimes}(b,c) (see the SI for corresponding movies). For a smooth yielding, we find in Fig.~\ref{fig:simulation_regimes}(b) that the non-affine displacements gradually fill the box as $\gamma$ increases, and concomitantly the stress displays an overshoot, as recently explored~\cite{shrivastav2016yielding,ghosh2017direct}. For the discontinuous case, the sharp stress drop corresponds to the sudden emergence of a system spanning shear band. By contrast with earlier work on shear-banding materials~\cite{shi2005strain,hassani2016localized}, the shear band in Fig.~\ref{fig:simulation_regimes}(c) appears suddenly in a single infinitesimal strain increment and does not result from the accumulation of many stress drops at large deformation. For an  intermediate regime between the discontinuous and continuous yielding ($T_{\rm ini} \approx 0.1$), strong sample-to-sample fluctuations are observed. Some samples show a sharp discontinuous yielding with a conspicuous shear band (similar to Fig.~\ref{fig:simulation_regimes}(c)), whereas other samples show smooth yielding with rather homogeneous deformation (similar to Fig.~\ref{fig:simulation_regimes}(b)). Such large sample-to-sample fluctuations are typical for systems with random critical points.

\section{The random critical point}

\label{criticality}

Having identified a regime where yielding takes place through a first-order discontinuity and a regime where it is a smooth crossover, we now provide quantitative support for the existence of a critical point separating them, as one would indeed expect on general grounds. The mean-field theory presented above supports this scenario and suggests that the critical point is in the universality class of an Ising model in a random field. This criticality should not be confused with the marginality predicted to be present in sheared amorphous solids irrespective of the degree of annealing and of the value of the strain~\cite{muller2015marginal,lin2015criticality}. This issue is discussed separately below and in the SI.  

\begin{figure*}[t]
\centering
\includegraphics[width=1.\linewidth]{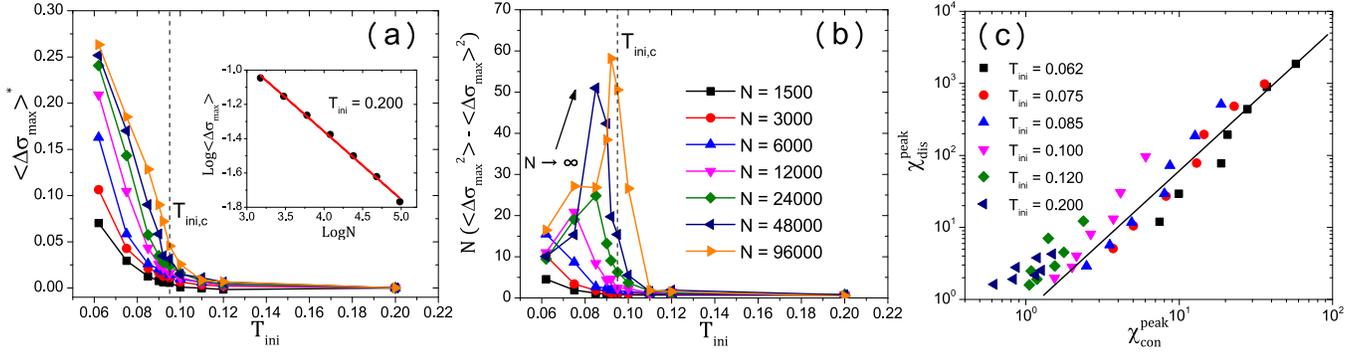}
\caption{Mean (a) and variance (b) of $\Delta \sigma_{\rm max}$ versus the preparation temperature $T_{\rm ini}$ for several system sizes $N$. In (a) we plot $\langle \Delta \sigma_{\rm max} \rangle^* \equiv \langle \Delta \sigma_{\rm max} \rangle-\langle \Delta \sigma_{\rm max} \rangle |_{T_{\rm ini}=0.2}$ to subtract the trivial high-temperature dependence which vanishes at large $N$, as shown in the inset. The critical temperature $T_{\rm ini, c}\approx 0.095$ is determined from the vanishing of the order parameter $\langle \Delta \sigma_{\rm max} \rangle^*$ in (a) and the growth of the maximum of the susceptibility $N(\langle \Delta \sigma_{\rm max}^2 \rangle- \langle \Delta \sigma_{\rm max} \rangle^2)$ in (b).
(c): Parametric plot of the connected and disconnected susceptibilities for all system sizes and several preparation temperatures. The straight line corresponds to the scaling $\chi_{\rm dis} \propto \chi_{\rm con}^2$, as found in the RFIM.}
\label{fig:critical}
\end{figure*}

As shown in Fig.~\ref{fig_MF}, the order parameter distinguishing the two regimes of yielding is the macroscopic stress drop. In the simulations, we measure its evolution by recording for each sample the maximum stress drop $\Delta\sigma_{\rm max}$ observed in the strain window $\gamma \in [0, 0.3]$. We have measured the mean value $\langle \Delta\sigma_{\rm max} \rangle$ as a function of the preparation temperature $T_{\rm ini}$ for several system sizes $N$. At the largest temperature, no macroscopic stress drop exists:  $\langle \Delta\sigma_{\rm max} \rangle$ simply reflects stress drops along the plastic branch and vanishes as $\langle \Delta\sigma_{\rm max} \rangle|_{T_{\rm ini}=0.2} \sim N^{-0.4}$, as shown in the inset of Fig.~\ref{fig:critical}(a). In the main panel of Fig.~\ref{fig:critical}(a), we subtract this trivial behavior from $\langle \Delta \sigma_{\rm max} \rangle$. We find that the maximum stress drop is zero above $T_{\rm ini} \approx 0.1$, and nonzero for lower temperatures. The system-size dependence confirms that this temperature evolution becomes crisp in the large-$N$ limit, and we locate the critical point at $T_{{\rm ini},c} \approx 0.095$. Complementary information is provided by studying the fluctuations of the maximum stress drop, which can be quantified through their  variance $N(\langle \Delta \sigma_{\rm max}^2 \rangle- \langle \Delta \sigma_{\rm max} \rangle^2)$ (not to be confused with the disconnected susceptibility $\chi_{\rm dis}=N(\langle \sigma^2 \rangle - \langle \sigma \rangle^2)$), shown in Fig.~\ref{fig:critical}(b). One finds that the variance goes through a maximum that increases with system size around $T_{{\rm ini},c} \approx 0.095$.


These results provide strong evidence of a critical point separating ductile from brittle behavior, with the mean stress drop $\langle \Delta \sigma_{\rm max} \rangle$ playing the role of an order parameter. Additional support comes from the study of the overlap function $q$ introduced in Ref.~\cite{jaiswal2016mechanical}. We find that the finite-size analysis of $q$ and of the overlap jump $\Delta q_{\rm max}$ at yielding follows the same pattern as $\sigma$ and $\Delta\sigma_{\rm max}$ qualitatively. This points toward a macroscopic discontinuity for well-annealed glasses and a mere crossover for poorly annealed cases (see the SI). Contrary to what found in Ref.~\cite{procaccia2017mechanical}, we find that the first-order transition behavior terminates at a temperature $T_{{\rm ini},c}$ above which a smooth stress overshoot is instead observed. 

Our findings are also corroborated by the analysis of the criticality of the sheared glass. As previously shown~\cite{karmakar2010statistical,lin2015criticality} an amorphous material quasi-statically sheared at zero temperature is marginal at all values of the strain. The physical reason is the presence of a pseudo-gap in the density of elementary excitation~\cite{muller2015marginal} which is characterized by a critical exponent $\theta >0$. This criticality implies a scale-free distribution of avalanche sizes, and by carefully analyzing the stress-drop statistics we have extracted the exponent $\theta$ as a function of $\gamma$ and $T_{\rm ini}$. As shown in the SI, we find that the discontinuous transition is associated with a discontinuous variation of $\theta$ and that large fluctuations of the stress drops associated with criticality generate a rapid change in $\theta$ versus $\gamma$ with the presence of a large maximum  for temperatures $T_{\rm ini}$ close to the critical point.

Our data do not allow us to measure the critical exponents associated to the RFIM critical point in a robust way. Yet it is possible to obtain a strong indication that the critical point and the first-order transition are governed by the universality class of an Ising model in a random field. In this case indeed the presence of quenched disorder leads to two distinct susceptibilities, $\chi_{\rm con}$ and $\chi_{\rm dis}$. A key signature of the presence of random-field disorder is that $\chi_{\rm dis} \propto \chi_{\rm con}^2$.
This scaling relation, which is exact in the mean-field limit, is valid in finite dimensions at the first-order transition and is also approximately verified by the conventional RFIM at the critical point~\cite{nattermann1998theory}. It indicates that disorder-induced sample-to-sample fluctuations provide the dominant source of fluctuations.
By looking at the parametric plot of the maximum of $\chi_{\rm dis}$ versus the maximum of $\chi_{\rm con}$ which is shown in Fig.~\ref{fig:critical}(c) for all system sizes and several preparation temperatures, one finds that the relation is indeed observed in our simulations, at least at and below a temperature $T_{\rm ini}=0.100 \gtrsim T_{{\rm ini},c}$ and for large $N$.

\section{Discussion and Conclusion}

Our analysis shows that irrespective of the nature of the amorphous material, yielding can come with two qualitatively different types of behavior, corresponding either to a discontinuous transition or to a smooth crossover. The transition between these two regimes occurs at a random critical point related to the RFIM, which naturally explains the  large sample-to-sample fluctuations observed in simulations. The type of yielding that a given material displays depends on its degree of annealing, a mechanism that differs dramatically from the processes at play in crystalline solids~\cite{rice1974}. Conceptually, increasing the annealing for a given particle interaction implies that the initial amorphous configuration is drawn from a deeper location of the glassy energy landscape, in which the local environments fluctuate less (lower disorder in the RFIM analogy). In practice, the degree of annealing can be tuned for some materials such as metallic and molecular glasses~\cite{shen2007plasticity,kumar2013critical,choi2013nanoscale,patrick17}, but would be more difficult to vary for others like emulsions and wet foams. Our approach shows that, given the particle size (for colloids), the preparation protocol (for emulsions), the cooling rate (for metallic glasses), a given amorphous material must belong to either one of the two yielding regimes. We suggest that colloids with a well-chosen range of particle sizes could be used to experimentally probe the random critical point separating the two yielding regimes. 

Our work is focused on the two possible yielding scenarios rather than on the stationary state reached at large deformation. In ductile glasses, one expects a stationary state independent of the initial condition as shear transformations are quickly healed so that plasticity can spread homogeneously. In the materials we dubbed ``brittle'' in the present work, large deformations would trigger cracks or shear bands that may remain well-localized in the sample (as we indeed find numerically). Our study does not allow us to study the propagation of the cracks themselves. 

There are several directions worth further studies to extend our results. On the theoretical side
it is important to introduce nonlocal elastic interactions mediated by an Eshelby-like  kernel in the proposed framework of an effective random-field Ising theory, which could potentially yield anisotropic avalanches that are not described by the traditional RFIM. 
This is essential to describe the role of nonperturbative and non-mean-field effects that have been argued to be important for the spinodal behavior of disordered finite-dimensional systems at zero temperature~\cite{nandi}. These correspond physically to rare regions that are able to trigger the failure in the material and are related to the shear bands found in simulations. We present numerical evidence already supporting this scenario in the SI (see also~\cite{wyart18}).  On the simulation side, it is interesting to study how the rheological setting affects the yielding scenario proposed in this work. Considering uniaxial tension or compression tests would be useful for a further detailed comparison between simulations and experiments. In addition, investigating the influence of a finite temperature and/or a finite strain rate on the simple situation studied here would also be a worthwhile extension.
Finally, one would like to understand better how the evolution of ductility with the initial disorder impacts the deformation and failure of glasses at larger length scales and make a connection with studies of macroscopic fracture in glasses. Because controlling ductility in amorphous solids is desirable for practical applications~\cite{schuh2007mechanical,schroers2004ductile}, our theoretical studies will hopefully lead to design-principle of more ductile glassy materials.  

\acknow{We thank H. Ikeda, F. Landes, A. Nicolas, A Ninarello, I. Procaccia, G. Tsekenis, P. Urbani, M. Wyart, F. Zamponi for helpful discussions. 
We would like to thank A. Ninarello for sharing very low temperature equilibrium configurations.
This work is supported by a grant from the Simons Foundation (No. 454933, LB, No. 454935, GB).}

\showacknow

\newpage

\appendix

{\bf{SUPPLEMENTAL INFORMATION}}

\section{Mean-field model of yielding}

\noindent {\it Derivation of the mean-field equations}

In order to obtain the mean-field equation on $P_\gamma(x)$ let us start from a mechanically stable state such that all mesoscopic blocks have a local stress $\sigma_i$ below the threshold value  $\sigma^{\rm th}$ that we set equal to one. It is useful to introduce the distance from threshold $x_i=\sigma^{\rm th}-\sigma_i=1-\sigma_i$. When a $x_i$ vanishes the block becomes unstable and jumps at a new value $x$, all the other blocks become closer to instability by the same amount $\frac{\mu_2}{N_{\rm b}(\mu_1+\mu_2)} x$.
Changing $\gamma$ in $\gamma+\delta \gamma$ shifts down all $x_i$'s by an amount $ 2 \mu_2 \delta \gamma$. In consequence, a fraction $ 2 \mu_2 \delta\gamma P_\gamma(0)$ of them become negative and undergo plastic reorganization~\footnote{From now on we neglect all sub-leading corrections in $1/N_{\rm b}$.},  which leads, independently for each block, to a new value $x$ drawn from the probability density $g(x)$. The corresponding change of $P_\gamma(x)$ is therefore 
\begin{equation}
dP_\gamma(x)= 2 \mu_2 \delta \gamma\left [ \frac{\partial P_\gamma(x)}{\partial x} + P_\gamma(0) g(x) \right ] \, .
\end{equation}
However, this is not the only contribution since due to the reinsertion in the pool of a fraction $P_\gamma(0)\delta\gamma$ of $x_i$'s, the mean-field interaction shifts down again all $x_i$ by an amount $\frac{\mu_2}{\mu_1+\mu_2}\bar x P_\gamma(0)\delta\gamma$, where $\bar x=\int_0^\infty dx\, x g(x)$, and gives rise to a contribution, akin to the one considered above, but with $\delta \gamma$ now replaced by $x_c P_\gamma(0)\delta\gamma$ and $x_c=\frac{\mu_2}{\mu_1+\mu_2} \bar x$. This process is iterated and leads to an infinite number of contributions: 
\begin{equation}
dP_\gamma(x)= 2 \mu_2 \delta \gamma\sum_{n=0}^\infty \left( x_c P_\gamma(0)\right)^n\left [ \frac{\partial P_\gamma(x)}{\partial x} + P_\gamma(0) g(x) \right ].
\end{equation}
By summing up the series, and dividing by $\delta\gamma$, one obtains the equation reported in the main text, which is an improved version of the equation studied in Ref.~\cite{jagla2014viscoelastic}. \\

\noindent {\it General analysis}

We now show that the mean-field equations naturally lead to the yielding regimes discussed in the main text (see also the explicit solution below).

Using the equations defining $\sigma_i$ and $x_i$ and taking their average, one obtains
\begin{equation}
\label{seq}
 \langle \sigma \rangle =1-\langle x\rangle =1-\int_0^{\infty} d x\, x \, P_\gamma(x)   \,.
\end{equation}
Multiplying the equation on $P_\gamma(x)$ by $x$, integrating over $x$ and inserting the resulting expression obtained
for $\frac{d\langle x\rangle}{d\gamma}$ in Eq.~(\ref{seq}) leads to 
\begin{equation}
    \label{P0}
    \frac{d \langle \sigma \rangle}{d \gamma} = 2 \mu_2 \frac{1-\bar x P_\gamma(0)}{1- x_c P_\gamma(0)} .
\end{equation}

The analysis of this equation points to two special values of $P_\gamma(0)$: $P_\gamma(0)= 1/\bar x$
at which $\frac{d \langle \sigma \rangle}{d\gamma}=0$ so that the steady state is reached and $P_\gamma(0)= 1/x_c$
at which $\frac{d \langle \sigma \rangle}{d\gamma}$ diverges. 


\begin{figure}
\begin{center}
\includegraphics[width=0.95\columnwidth]{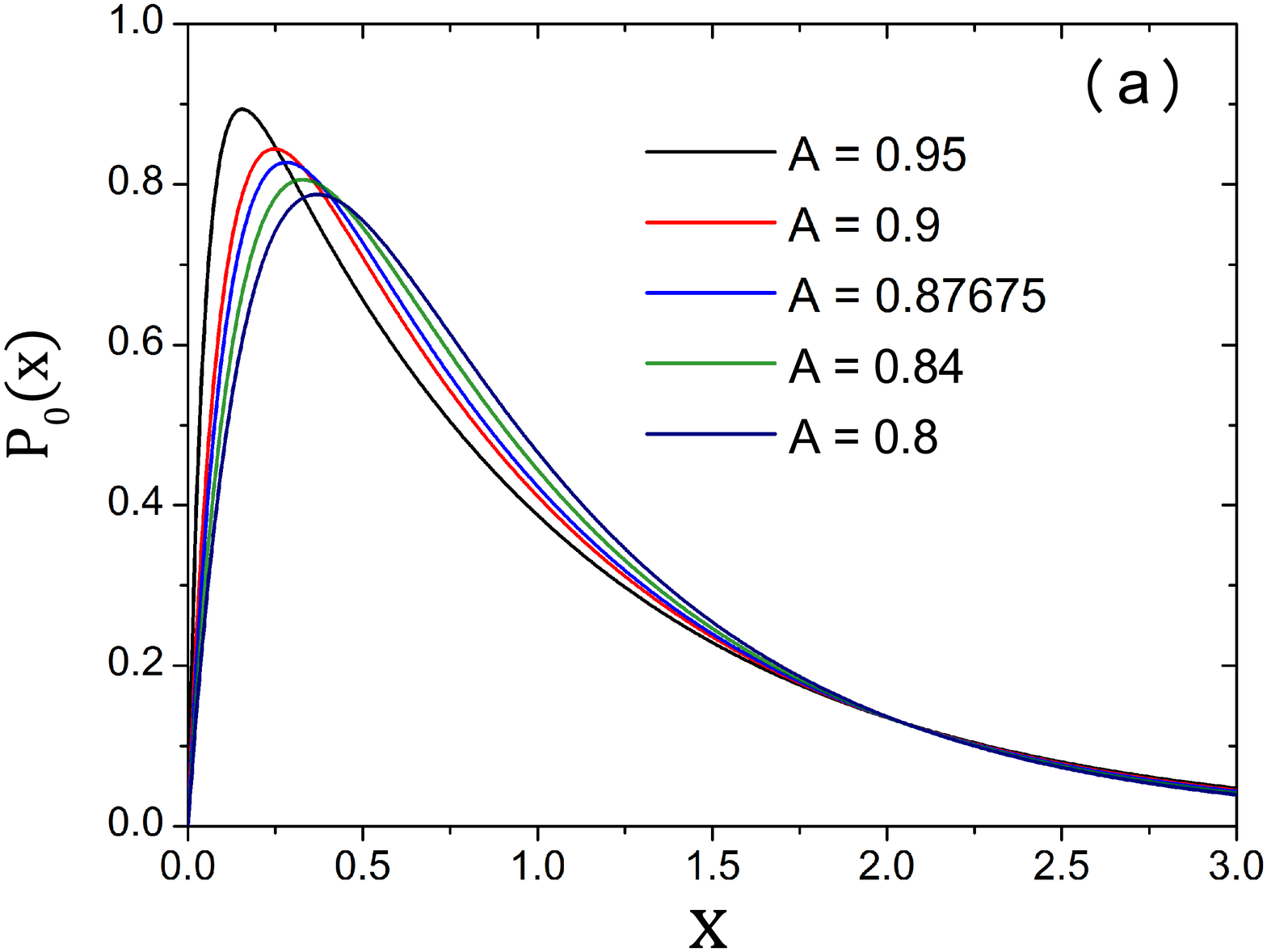}
\includegraphics[width=0.95\columnwidth]{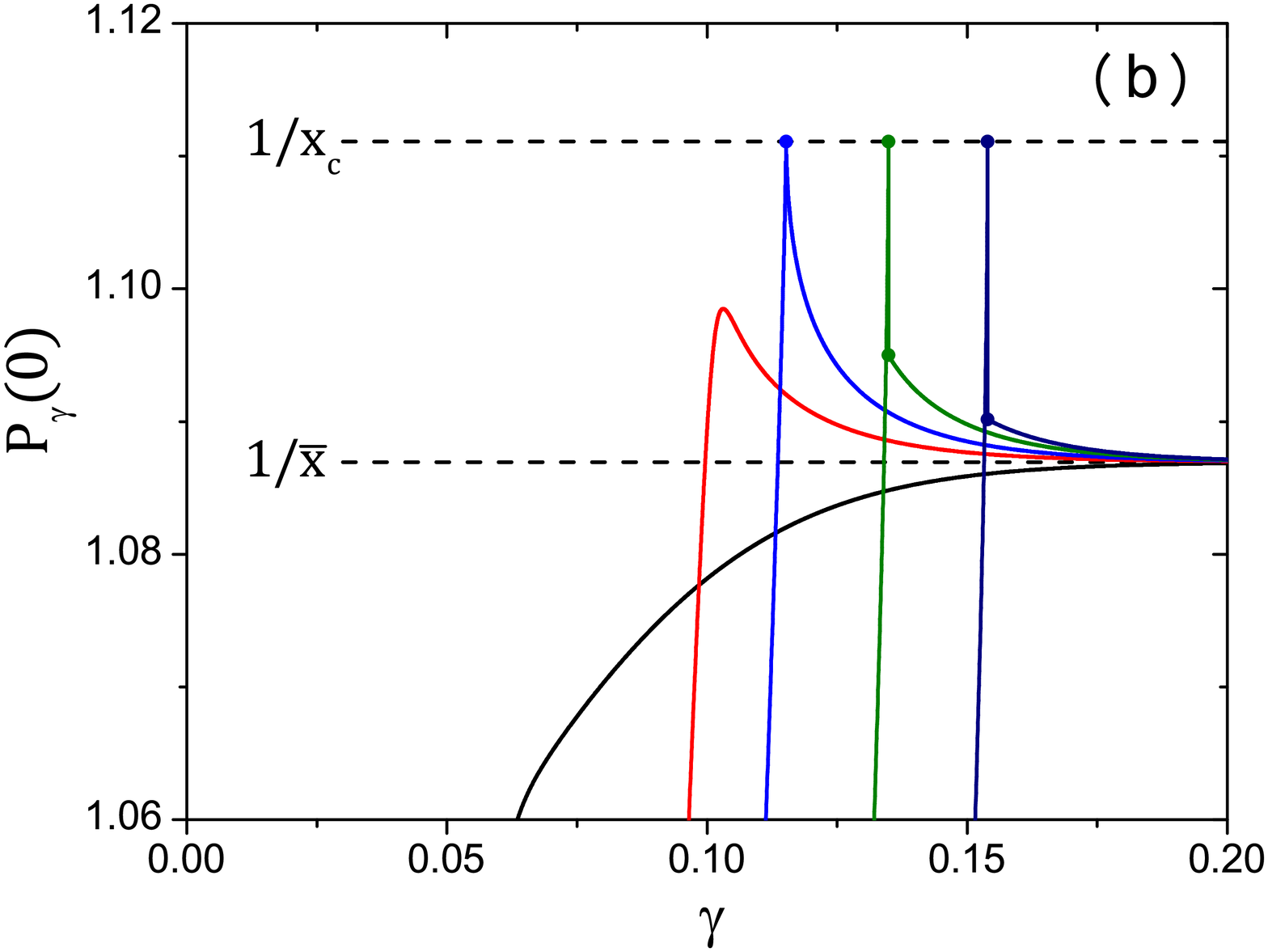}
\caption{
Effect of the degree of annealing described by the parameter $A$ in Eq.~(\ref{initial}):
(i) discontinuous regime (dark blue, $A=0.8$, and green, $A=0.84$), (ii) continuous regime with overshoot (red, $A=0.9$) and (iii) monotonic regime (black , $A=0.95$). The the critical point (blue) is identified at $A_c=0.87675$.
(a) The initial distribution $P_{\gamma=0}(x)$ and (b) the evolution of $P_{\gamma}(x=0)$ are shown in the different regimes.}
\label{fig:SIMF}
\end{center}
\end{figure}  
\noindent

We have now all we need to discuss the existence of three regimes depending on the initial distribution $P_{\gamma=0}(x)$. In the following we discuss our results, that hold for generic $g(x)$'s, and illustrate them in the special case already discussed in the main text.  
\begin{itemize}
\item {\it Monotonic yielding curve.} This regime corresponds to a $P_\gamma(0)$ that remains always below $1/\bar x$, value reached   monotonically only for $\gamma\rightarrow \infty$ (\ref{fig:SIMF}). This case corresponds to the largest $A=0.95$ (black line) in Figs. 1 (main text)  and \ref{fig:SIMF} in which $\langle \sigma \rangle$ increases monotonically towards its asymptotic value.
\item {\it Overshoot.} In this case $P_\gamma(0)$ is small for small strain, increases and crosses the value $1/\bar x$ at a finite value $\gamma=\gamma_{\rm max}$. By increasing $\gamma$ more it reaches a maximum, whose value is smaller than $1/x_c$, and then it starts decreasing and asymptotically converges from above to $1/\bar x$. In this case, the curve  $\langle \sigma \rangle$ versus $\gamma$ displays a maximum at $\gamma=\gamma_{\rm max}$ but no singular behavior since $\frac{d\langle \sigma\rangle}{d\gamma}$ remains bounded ($A=0.9$ (red line) in Figs.~1 (main text) and \ref{fig:SIMF}).
\item {\it Discontinuous yielding.} This is similar to the previous case except that $P_\gamma(0)$ keeps increasing after $\gamma_{\max}$  and eventually reaches the value $1/x_c$ for $\gamma=\gamma_{Y}$. This means that the curve  $\langle \sigma \rangle$ versus $\gamma$ first goes though a maximum and then has an infinite negative derivative at $\gamma=\gamma_{Y}$ ($A=0.8$ and $0.84$ (dark blue and green line, respectively) in Figs. 1 (main text)  and \ref{fig:SIMF}). At this point, which can be considered as a {\it spinodal point}, the distribution $P_\gamma(x)$ has a discontinuous transition due to the divergence of the right-hand side of its evolution equation. Due to Eq.~(\ref{seq}), this also leads to a discontinuity in the macroscopic stress $\langle \sigma \rangle$.
The behavior of $P_\gamma(0)$ can be understood by studying its evolution equation, 
\begin{equation}\label{eqpx2}
\frac{\partial P_\gamma(0)}{\partial \gamma}= \frac{2 \mu_2}{1-x_c P_\gamma(0)} \left [ \frac{\partial P_\gamma(0)}{\partial x} + P_\gamma(0) g(0) \right ]\,.
\end{equation}  
At $\gamma=\gamma_Y^-$ its right-hand side is positively divergent since generically $\left [ \frac{\partial P_\gamma(0)}{\partial x} + P_\gamma(0) g(0) \right ]>0$. Then, a macroscopic avalanche takes place. At the same time, $P_\gamma(0)$ jumps to a value that is less than $1/x_c$, and at $\gamma=\gamma_Y^+$ the right-hand side of Eq.~(\ref{eqpx2}) is no longer singular. When approaching the spinodal point, for $\gamma \to \gamma_Y^-$, one expects that $P_\gamma(0)\approx 1/x_c - A(\gamma_Y-\gamma)^{\alpha_{-}}$. Inserting this ansatz in the equation for $ P_\gamma(0)$  one finds $\alpha_-=1/2$. On the other hand for $\gamma \to \gamma_Y^+$ the behavior is regular. This in turn implies
\begin{eqnarray}
 \langle \sigma \rangle- \sigma_{\rm sp}&\propto& (\gamma_Y-\gamma)^{1/2},  \,\,\,\,\gamma<\gamma_Y, \\
\sigma_+ -\langle \sigma \rangle &\propto& (\gamma-\gamma_Y), \qquad \gamma>\gamma_Y,
\end{eqnarray}
where $\sigma_{\rm sp}=\langle \sigma \rangle ( \gamma_Y^- ) > \sigma_+ = \langle \sigma \rangle( \gamma_Y^+) $ due to the discontinuity. As shown in Fig.~1 (main text) this corresponds to a singular behavior with a square-root singularity before the discontinuity of $\langle \sigma \rangle$ and a regular behavior after. 
\end{itemize}
We now focus on the {\it critical point} separating the overshoot and discontinuous regimes.  It takes place when, for $\gamma=\gamma_{Y_c}$, both  $\left [ \frac{\partial P_\gamma(0)}{\partial x} + P_\gamma(0) g(0) \right ]=0$ and $P_\gamma(0)=1/x_c$ are satisfied.  In this case the right-hand side of Eq.~(\ref{eqpx2}) does not diverge anymore. To study the critical point, it is useful to focus on the evolution equation of 
$ \frac{\partial P_\gamma(0)}{\partial x}$:
\begin{equation}\label{eqpx3}
\frac{\partial }{\partial \gamma}\frac{\partial P_\gamma(0)}{\partial x}= 
 \frac{2 \mu_2}{1-x_c P_\gamma(0)} \left [ \frac{\partial^2 P_\gamma(0)}{\partial x^2} + P_\gamma(0) g'(0) \right ]\,.
\end{equation}  
Inserting in Eq. (\ref{eqpx3}) the ansatz   $P_\gamma(0)\approx 1/x_c + A|\gamma_{Y_c}-\gamma|^{\alpha}$ and assuming that the term within the square brackets is generically different from zero for $\gamma =\gamma_{Y_c}$, one finds that $ \frac{\partial P_\gamma(0)}{\partial x} + P_\gamma(0)g(0) \propto \text{sgn}(\gamma_{Y_c}-\gamma)|\gamma_{Y_c}-\gamma|^{1-\alpha}$. Using this result in Eq.~(\ref{eqpx2}) one finally obtains $\alpha=2/3$. 
Due to Eq. (\ref{seq}), this leads to a critical behavior of $\langle \sigma \rangle$ for $\gamma$  near $\gamma_{Y_c}$ with
\[
\ \langle \sigma \rangle \approx \sigma_{c}+B\,{\rm sgn}(\gamma_{Y_c}-\gamma)|\gamma_{Y_c}-\gamma|^{1/3}\,,
\]
where $B$ is a constant and $\sigma_c=\langle \sigma \rangle(\gamma_{Y_c})$. 

By using the results of Ref.~\cite{jagla2014viscoelastic} one can also determine the behavior of the avalanche size distribution,
\[
\mathcal P(S) \sim S^{-\tau} e^{-\frac{S}{2 S_{\rm cut}}}\,,
\]
where it is found that $\tau=3/2$ and $S_{\rm cut}=1/[1-x_cP_\gamma(0)]^2$. The latter relation leads to $S_{\rm cut}\sim (\gamma_Y-\gamma)^{-1}$ for $\gamma<\gamma_Y$ near the spinodal point of the discontinuous yielding and $S_{\rm cut}\sim \vert \gamma_{Y_c}-\gamma\vert^{-4/3}$ near the critical point.

These results for the scaling behavior near the spinodal point and near the critical point are identical to those obtained for the RFIM quasi-statically driven at zero temperature within the mean-field theory \cite{dahmen-sethna}. In the correspondence between yielding of an amorphous solid and hysteresis of the RFIM, the discontinuous yielding corresponds to the discontinuous jump of the magnetization that takes place for weak disorder, the continuous yielding with an overshoot is analog to the smooth behavior of the magnetization found for strong disorder, and the critical point at $\gamma=\gamma_{Y_c}$ is the counterpart of the critical point of the RFIM and corresponds to plain old criticality \cite{sethna}. Note that the case of a monotonic yielding  curve has no counterpart in the RFIM. \\

\noindent {\it Exact solution in the exponential case}

The above general analysis is supported by the explicit solution of the model for a specific choice of the distribution $g(x)$ and setting $2 \mu_2=1$. This solution is shown in Fig. 1 (main text) and Fig.~\ref{fig:SIMF}.

To obtain the solution of the mean-field equation for $P_\gamma(x)$ [Eq. (1) of the main text], it is convenient to introduce the auxiliary variable $y$ such that
\[
\frac{dy}{d\gamma}=\frac{1}{1-x_c P_\gamma(0)}\,.
\]
This variable plays the role of the so-called plastic strain in elasto-plastic models. One then easily derives that the solution satisfies
\begin{equation}
\label{eq:MFgeneral}
P_y(x)= P_0(x+y) +\int_x^{x+y} dy' g(y') P_{x+y-y'}(0)\, ,
\end{equation}
where $P_y(x)\equiv P_{\gamma(y)}(x)$ and
\begin{equation}
\gamma(y)=y-x_c \int_0^y dy' P_{y'}(0)\,.
\end{equation}
Rather than pursuing formal developments, we illustrate the solution for the special case where the distribution $g(x)$ is exponential:
\[
g(x)=\frac 1{\bar x}e^{- x / {\bar x}}\, .
\]
The solution of Eq.~(\ref{eq:MFgeneral}) is now easily found to be
\begin{equation}
\label{eq:MFexp}
P_y(x)= P_0(x+y) +\frac 1{\bar x}e^{-\frac x{\bar x}} \int_0^{y} dy' P_{0}(y')\,.
\end{equation}
One is more specifically interested by the macroscopic stress $\langle \sigma \rangle$ and $P_\gamma(0)$. For a given initial distribution $P_0(x)$, they can be obtained from parametric plots of the following equations:  
\begin{equation}
\begin{aligned}
\label{eq:MFexp_final}
&\gamma(y)=y-\frac{x_c}{\bar x} \left [\int_0^{y} dy' (y-y'+\bar x) P_{0}(y')\right ], \\&
\langle \sigma \rangle (y)=y\left [1-\int_0^ydy' P_0(y')\right ] + \int_0^y dy' (y'-\bar x) P_0(y') ,  \\&
P_y(0)= P_0(y) +\frac 1{\bar x} \int_0^{y} dy' P_{0}(y')\,.
\end{aligned}
\end{equation}
The above expressions are valid for any initial distribution $P_0(x)$ that satisfies some constraints (on top of normalization): (i) for $\gamma=0$ we have $\langle \sigma \rangle=0$, setting $\sigma^{\rm th}= 1$ this imposes that $\int_0^{\infty} dx x P_{0}(x)=1$,
(ii) the slope of $\langle \sigma \rangle$ versus $\gamma$ is positive at the origin, leading to $P_0(0)<1/\bar x$. Finally, note  that by construction $x_c<\bar x$, and from the physical requirement that $\langle \sigma \rangle(y\to \infty)>0$ one must have $\bar x<1$.

It is straightforward to show that the above solution behaves near the spinodal point and the critical point exactly as predicted by the preceding general analysis. Furthermore, to illustrate the outcome of the mean-field description of yielding in the main text we have chosen for $P_0(x)$ a combination of two exponential functions,
\begin{equation}
\label{initial}
P_{\gamma=0}(x)=\left(e^{-x/A}-e^{-x/(1-A)}\right)/(2A-1)\,,
\end{equation}
which has the merit of having only a single control parameter, $A$ (with $1/2<A<1$), which characterizes in the model the degree of annealing of the glass sample.

\section{Simulation methods}

We consider a three-dimensional atomistic model with a continuous size polydispersity, where the particle diameter $d$ of each particle is randomly drawn from the following particle-size distribution: 
$f(d) = Cd^{-3}$ for $d \in [ d_{\rm min}, d_{\rm max} ]$, where $C$ is a normalization constant. We choose a polydispersity parameter $\delta=\sqrt{\overline{d^2} - \overline{d}^2}/\overline{d}=0.23$, where $\overline{(\cdots)}=\int f(d) (\cdots)\mathrm{d}d$, with $d_{\rm min} / d_{\rm max} = 0.45$. We use $\overline{d}$ as the unit length. We simulate systems composed of $N$ particles in a cubic cell of volume $V$ with periodic boundary conditions~\cite{allen2017computer}. The following pairwise soft-sphere potential is used:
\[
v_{ij}(r) = v_0 \left( \frac{d_{ij}}{r} \right)^{12} + c_0 + c_1 \left( \frac{r}{d_{ij}} \right)^2 + c_2 \left( \frac{r}{d_{ij}} \right)^4, \label{eq:soft_v}
\]
with
\[
d_{ij} = \frac{(d_i + d_j)}{2} (1-0.2 |d_i - d_j|),
\]
where $v_0$ is the unit of energy. Nonadditivity of the diameters is introduced for convenience, as it prevents crystallization more efficiently and thus enhances the  glass-forming ability of the numerical model. The constants, $c_0$, $c_1$ and $c_2$, are chosen so that the first and second derivatives of $v_{ij}(r)$ become zero at the cut-off $r_{\rm cut}=1.25 d_{ij}$. We set the number density $\rho=N/V=1.0$.

We employ a swap Monte-Carlo method~\cite{ninarello2017models}. This approach is a very efficient thermalization algorithm which enables us to obtain very deep supercooled liquids. Details concerning these efficient simulations are provided in Ref.~\cite{ninarello2017models}. To perform the quench of the system down to zero temperature, we use a conjugate-gradient method~\cite{wright1999numerical}.

The athermal quasi-static shear method~\cite{maloney2006amorphous,rodney2011modeling} consists of a succession of tiny uniform shear deformations with $\Delta \gamma=10^{-4}$ followed by energy minimization via the conjugate-gradient method. We perform these simulations using Lees-Edwards boundary conditions~\cite{allen2017computer}.
We have varied $\Delta \gamma$ systematically from $\Delta \gamma=10^{-3}$ to $\Delta \gamma=3 \times 10^{-6}$ for some $N=12000$ samples at $T_{\rm ini}=0.062$ and found that below $\Delta \gamma=3 \times 10^{-4}$, the location of yielding hardly changes. Thus we conclude that $\Delta \gamma=10^{-4}$ is a good choice for our purpose.

To obtain the averaged value $\langle (\cdots) \rangle$ in simulations, we average over $800, 400, 200, 100, 100, 50$, and $25-50$ samples for $N=1500, 3000, 6000, 12000, 24000, 48000$, and $96000$ systems, respectively.

\section{Glass preparation}

\begin{figure}
\begin{center}
\includegraphics[width=0.95\columnwidth]{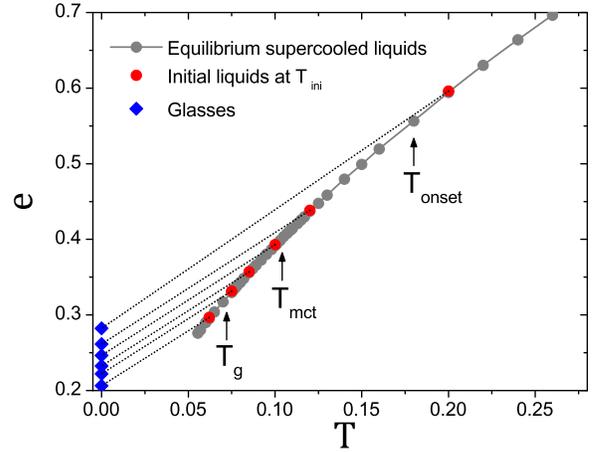}
\caption{
Potential energy of equilibrium supercooled liquids (circles) and associated glass states (diamonds). The dashed lines illustrate the rapid quench from the liquid states at $T_{\rm ini}$ to the corresponding glass states. The arrows denote representative reference temperatures: Onset of slow dynamics $T_{\rm onset}$, mode-coupling crossover $T_{\rm mct}$, and experimental glass-transition temperature estimated from the parabolic law~\cite{elmatad2009corresponding} (see the details in Ref.~\cite{ninarello2017models}). This plot illustrates the unprecedented range of glass stability achieved in the present work.}
\label{fig:EOS}
\end{center}
\end{figure}  

We summarize our glass preparation in the potential energy $e$ versus  temperature $T$ plot of Fig.~\ref{fig:EOS}. 
Thanks to the swap Monte Carlo simulation, we can equilibrate supercooled liquids down to, and even below, the estimated experimental glass transition temperature $T_{\rm g}=0.072$~\cite{ninarello2017models}. To prepare the glass samples to be sheared in athermal quasi-static conditions, we rapidly quench equilibrated supercooled liquid configurations down to zero temperature by using the conjugate-gradient method~\cite{wright1999numerical}.

\section{Non-affine displacements}

Here we explain the definition of the non-affine displacement used in the color bar of the snapshots (main text) and movies (SI). The position of the $i$-th particle at the strain $\gamma$ is
${\bf r}_i (\gamma)=(x_i(\gamma),y_i(\gamma),z_i(\gamma))$.
We introduce the modified position obtained by subtracting the displacement due to affine deformation:  
${\bf r}_i^{\rm NA} (\gamma)=(x_i^{\rm NA}(\gamma),y_i(\gamma),z_i(\gamma))$, where
\begin{equation}
x_i^{\rm NA}(\gamma) = x_i(\gamma) - \int_0^{\gamma} \mathrm{d}\gamma' y_i(\gamma'). 
\end{equation}
Trivially, we get ${\bf r}_i^{\rm NA} (0)={\bf r}_i (0)$. We then define the non-affine displacement as $|{\bf r}_i^{\rm NA} (\gamma)-{\bf r}_i(0)|$, which we use to detect mobile regions in snapshots and movies.

\section{Stress and overlap}

\noindent {\it Derivative of the stress}

\begin{figure}
\begin{center}
\includegraphics[width=0.95\columnwidth]{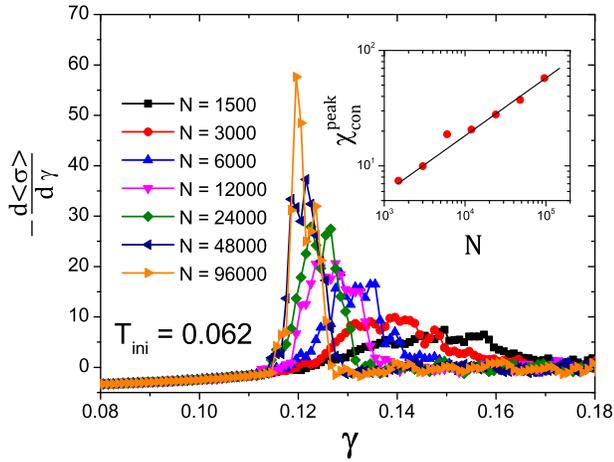}
\caption{
The derivative of $\langle \sigma \rangle$ with respect to $\gamma$ for $T_{\rm ini}=0.062$.
Inset: The divergence of the maximum of $\chi_{\rm con}$ is proportional to $\sqrt{N}$ shown with the straight line.
}
\label{fig:derivative_stress}
\end{center}
\end{figure}  

To compute $\chi_{\rm con}=-d \langle \sigma \rangle/ d \gamma$, direct numerical differentiation of $\langle \sigma \rangle$ with respect to $\gamma$ is too noisy as there are significant fluctuations between two successive data points separated by the small chosen interval $\Delta \gamma$. Thus, we first smooth the data by averaging over $10$ adjacent data points. We also perform the same smoothing procedure for the variance, $\chi_{\rm dis}=N(\langle \sigma^2 \rangle-\langle \sigma \rangle^2)$. We have checked that our conclusions do not change when choosing the number of adjacent data points in the range between $5$ and $20$. For each averaged data point $k$, we compute the derivative via $(d \langle \sigma \rangle/ d \gamma)_k=(\langle \sigma \rangle_{k+1}-\langle \sigma \rangle_{k-1})/(\gamma_{k+1}-\gamma_{k-1})$.
We show the result for the lowest preparation temperature in Fig.~\ref{fig:derivative_stress}. Similarly to the disconnected susceptibility $\chi_{\rm dis}$ shown in the main text, $\chi_{\rm con}$ has a peak and it steadily grows with increasing $N$, which supports the existence of a sharp first-order yielding transition in the thermodynamic limit. Note that $\chi_{\rm con}^{\rm peak}$ and $\chi_{\rm dis}^{\rm peak}$ increase as $\sqrt{N}$ and $N$, respectively as shown in the inset of Fig~\ref{fig:derivative_stress} and section 4 of the main text, which is a signature of a first-order transition in the presence of a random field.\\

\noindent {\it Analysis in terms of the overlap}

In the main text, we use the stress $\sigma$ to discuss the nature of yielding. A very similar conclusion is reached by using instead the overlap function. The collective overlap $q$ is defined as~\cite{jaiswal2016mechanical}:
\begin{equation}
q(\gamma) = \frac{1}{N} \sum_{i, j} \theta(a-|{\bf r}_i^{\rm NA}(\gamma)-{\bf r}_j (0)|), 
\label{eq:overlap}
\end{equation}
where $\theta(x)$ is the step function. We have set $a=0.23$. Note that we have also analyzed the ``self'' version of Eq.~(\ref{eq:overlap}) (where the sum is over a single particle), but the difference between the self and collective functions is found to be negligible. 

\begin{figure}
\begin{center}
\includegraphics[width=0.48\columnwidth]{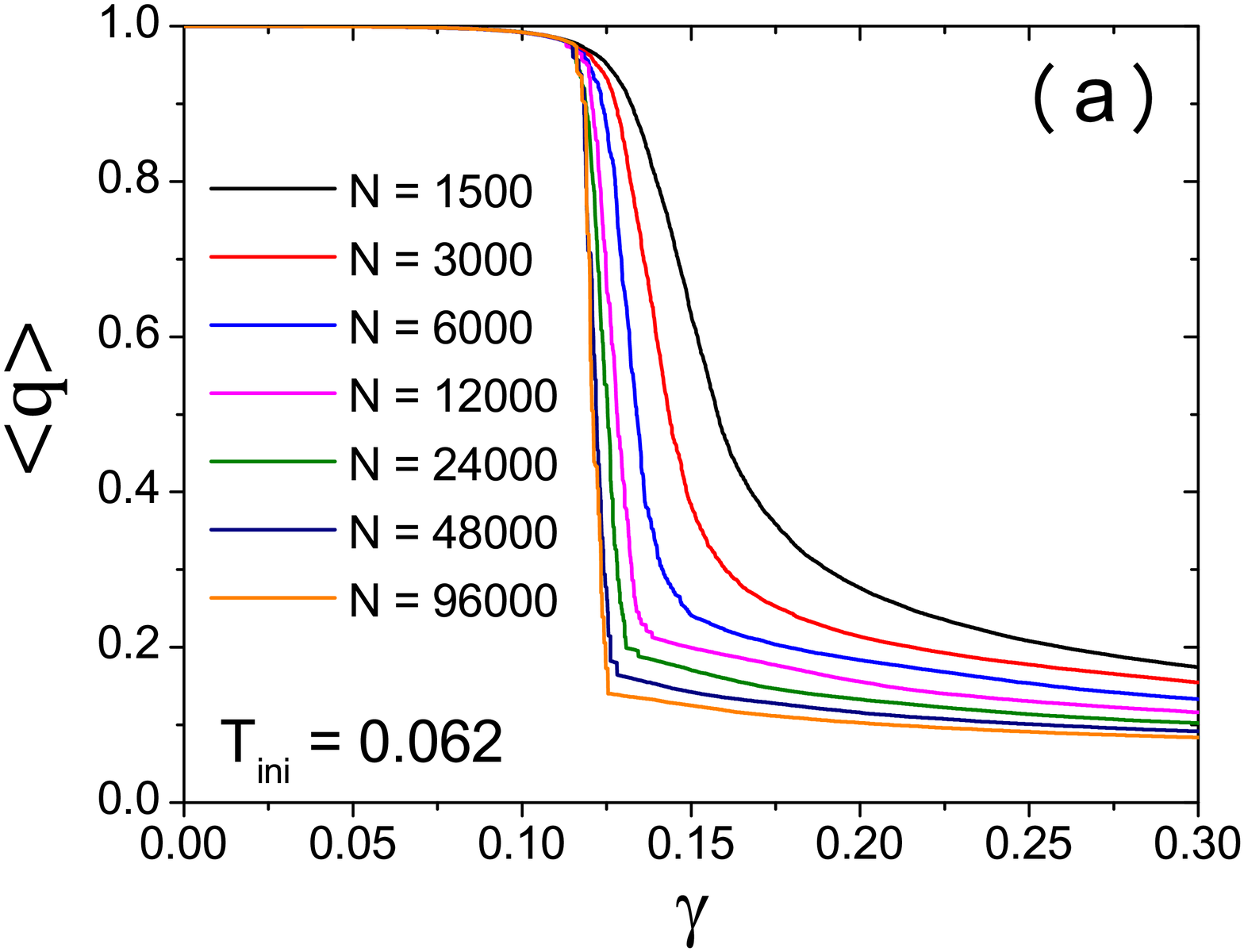}
\includegraphics[width=0.48\columnwidth]{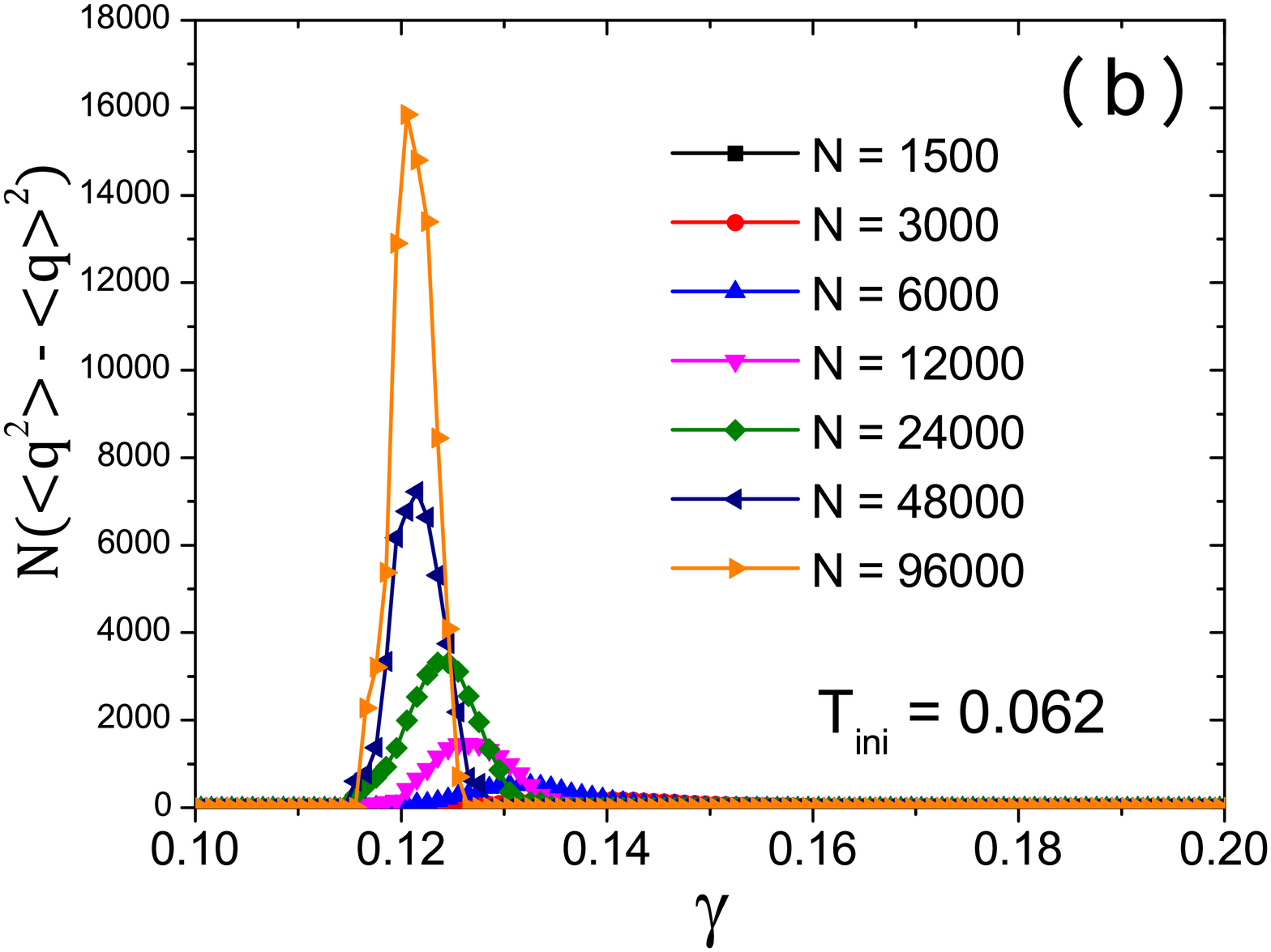}
\includegraphics[width=0.48\columnwidth]{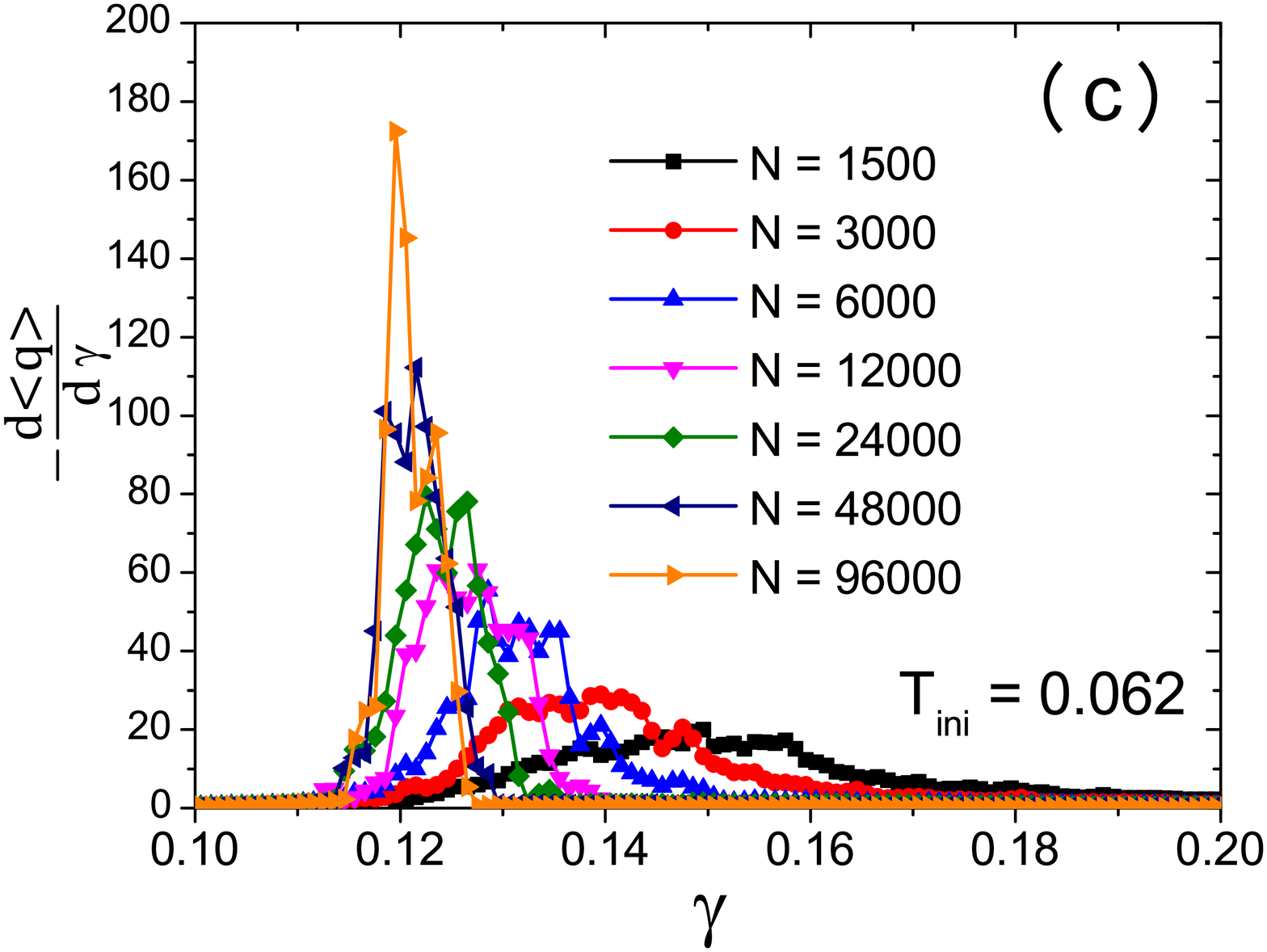}
\includegraphics[width=0.48\columnwidth]{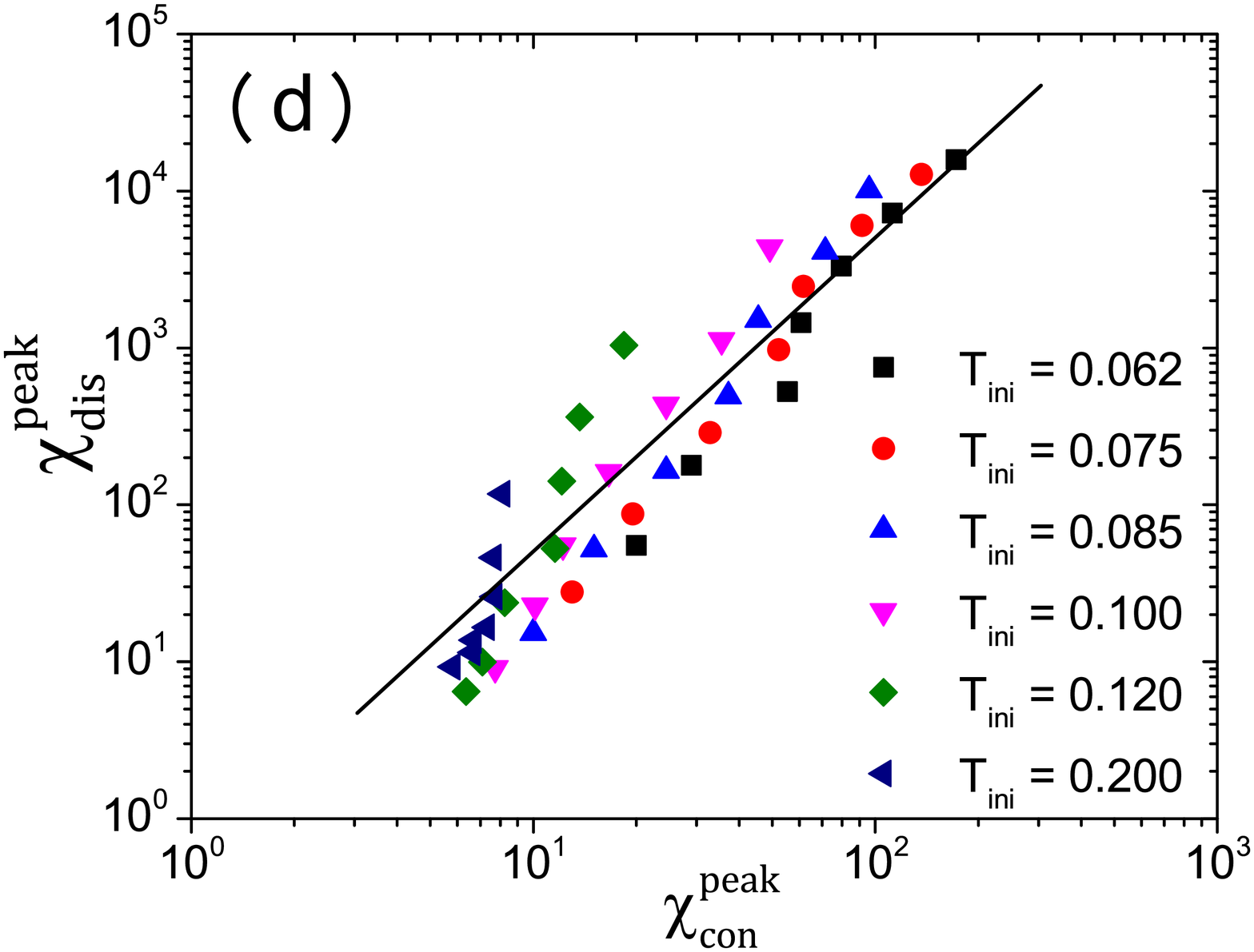}
\caption{
Analysis by the overlap function. Mean (a), variance (b), and derivative (c) of $q$ as a function of $\gamma$ for $T_{\rm ini}=0.062$.
(d): The peak values of $\chi_{\rm con}$ vs. $\chi_{\rm dis}$. The straight line corresponds to the expected behavior, $\chi_{\rm dis} \propto \chi_{\rm con}^2$, predicted from the random-field Ising model.
}
\label{fig:overlap}
\end{center}
\end{figure}  

As shown in Fig.~\ref{fig:overlap}, we observe a sharp drop of $\langle q \rangle$ (Fig.~\ref{fig:overlap}(a)) and the associated divergence of $\chi_{\rm dis}$  (Fig.~\ref{fig:overlap}(b)) and $\chi_{\rm con}$ (Fig.~\ref{fig:overlap}(c)). Finally, we also find that  $\chi_{\rm dis} \propto \chi_{\rm con}^2$ around and below $T_{{\rm ini},c}$ as predicted for an Ising model in a random field (Fig.~\ref{fig:overlap}(d)).

\begin{figure}
\begin{center}
\includegraphics[width=0.48\columnwidth]{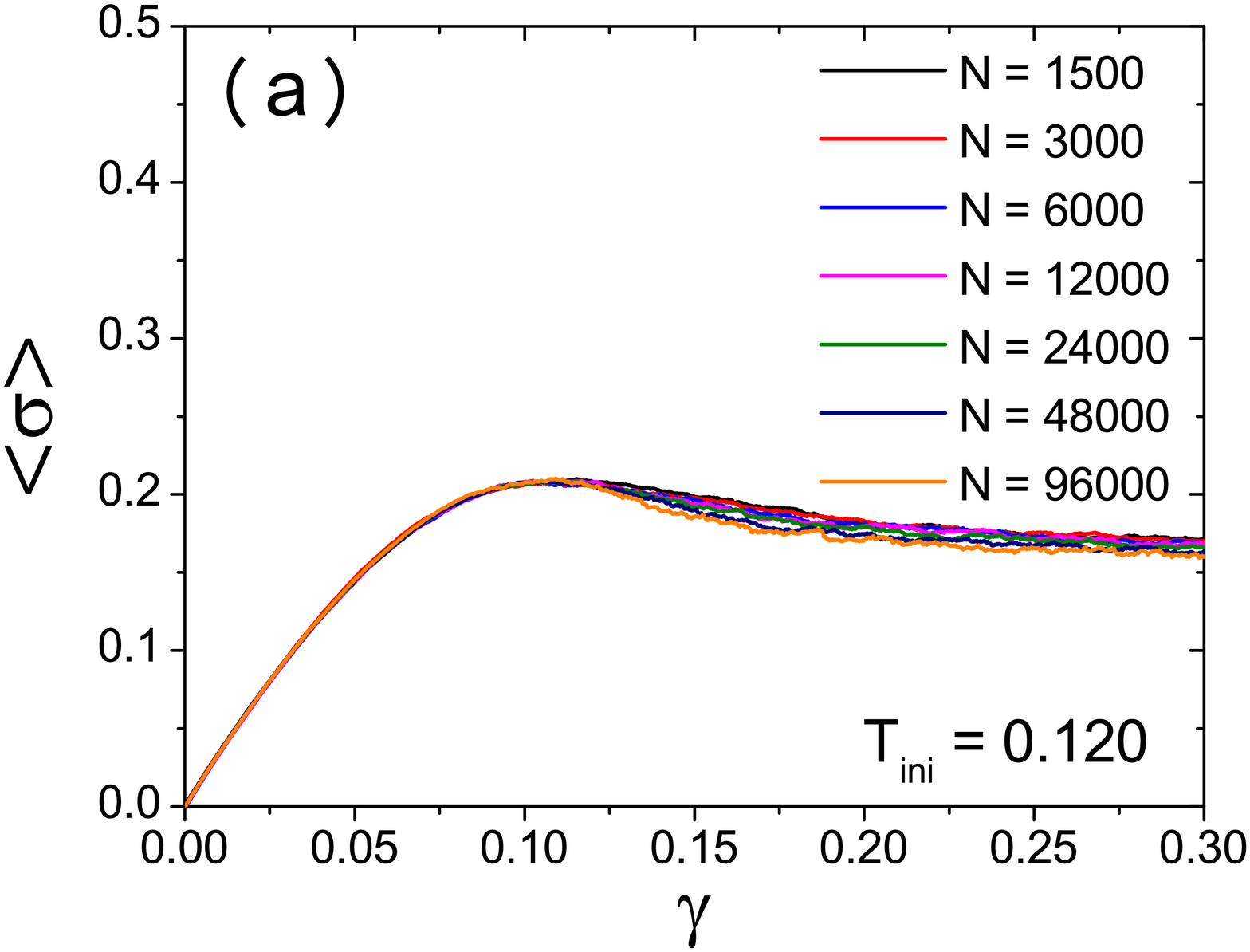}
\includegraphics[width=0.48\columnwidth]{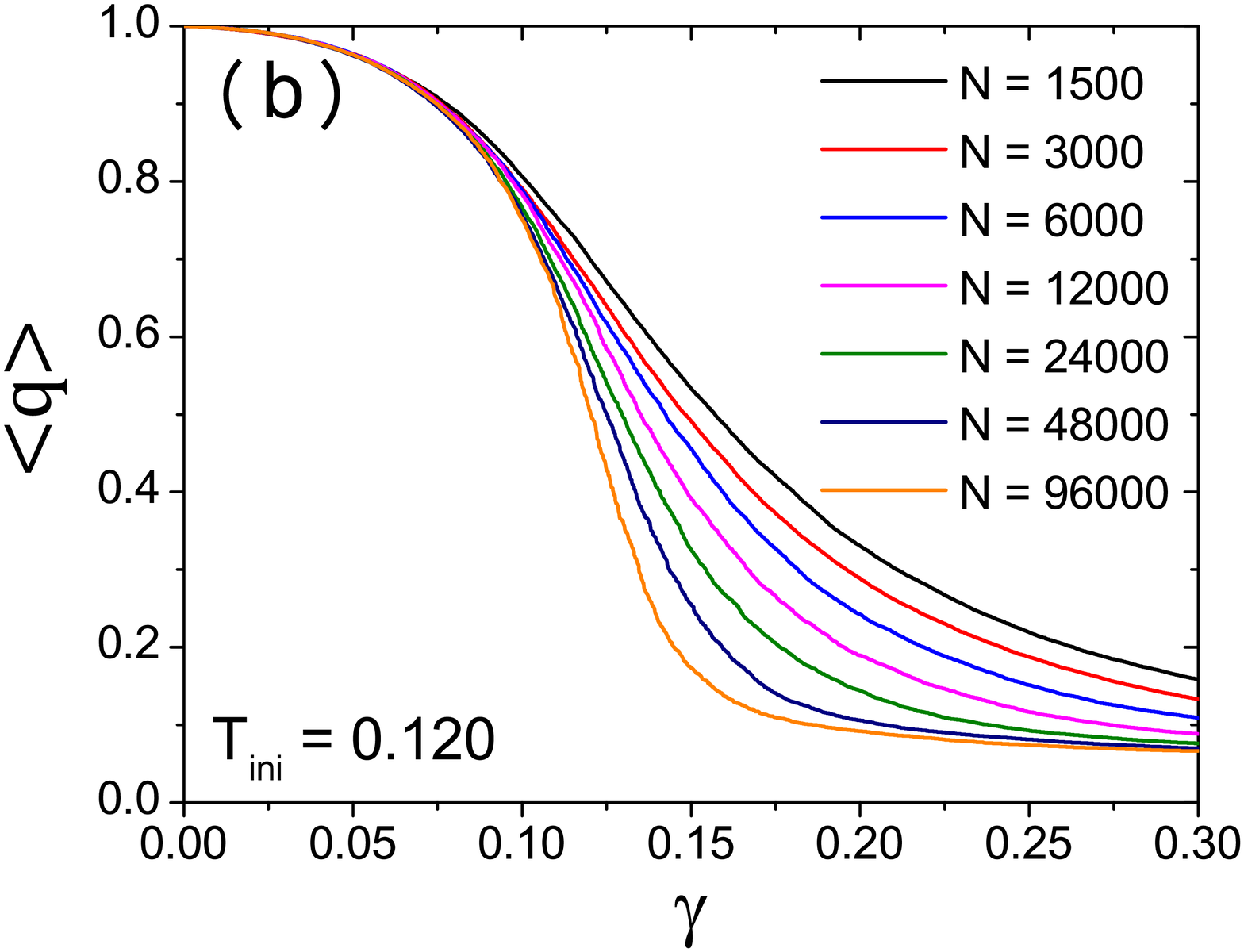}
\includegraphics[width=0.48\columnwidth]{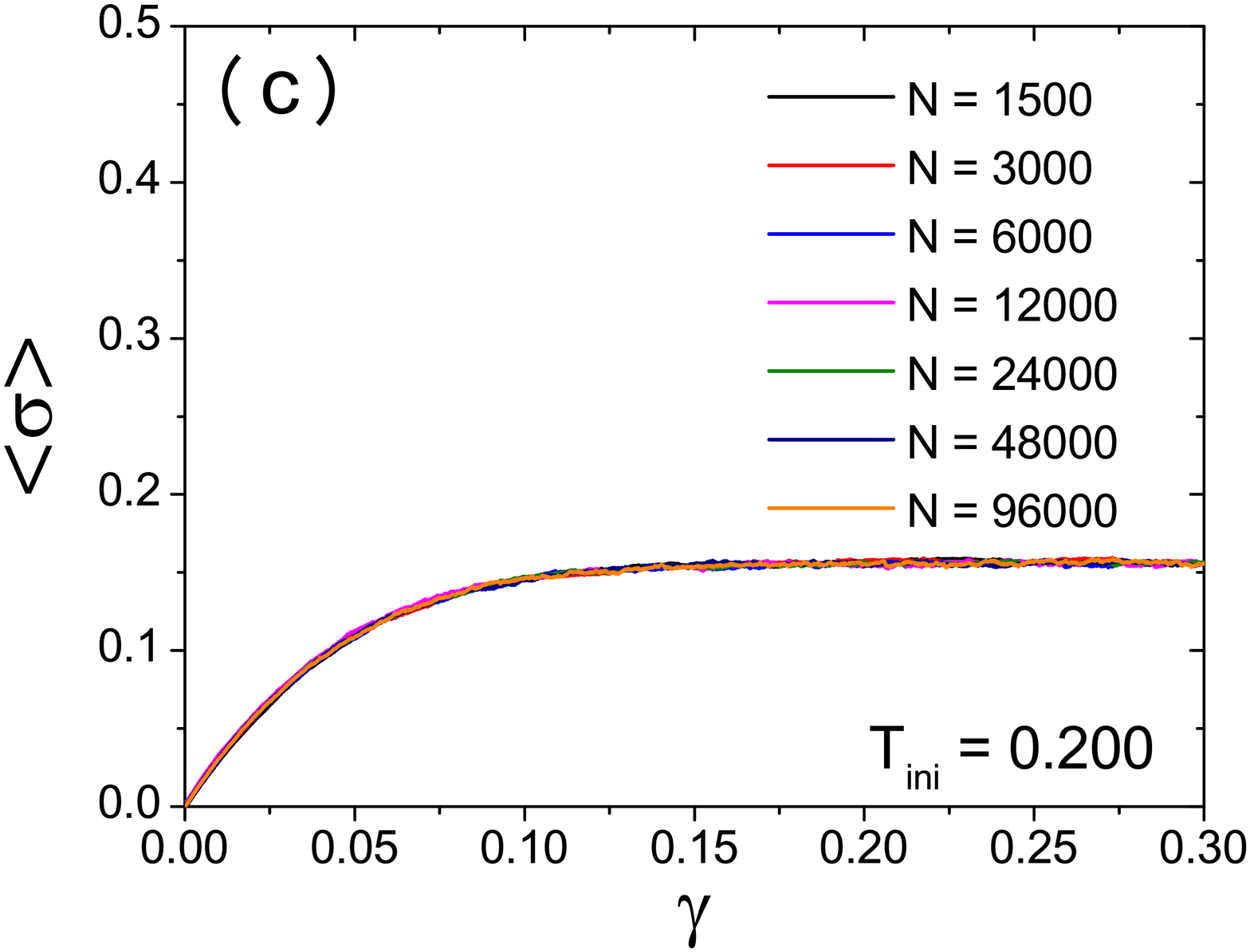}
\includegraphics[width=0.48\columnwidth]{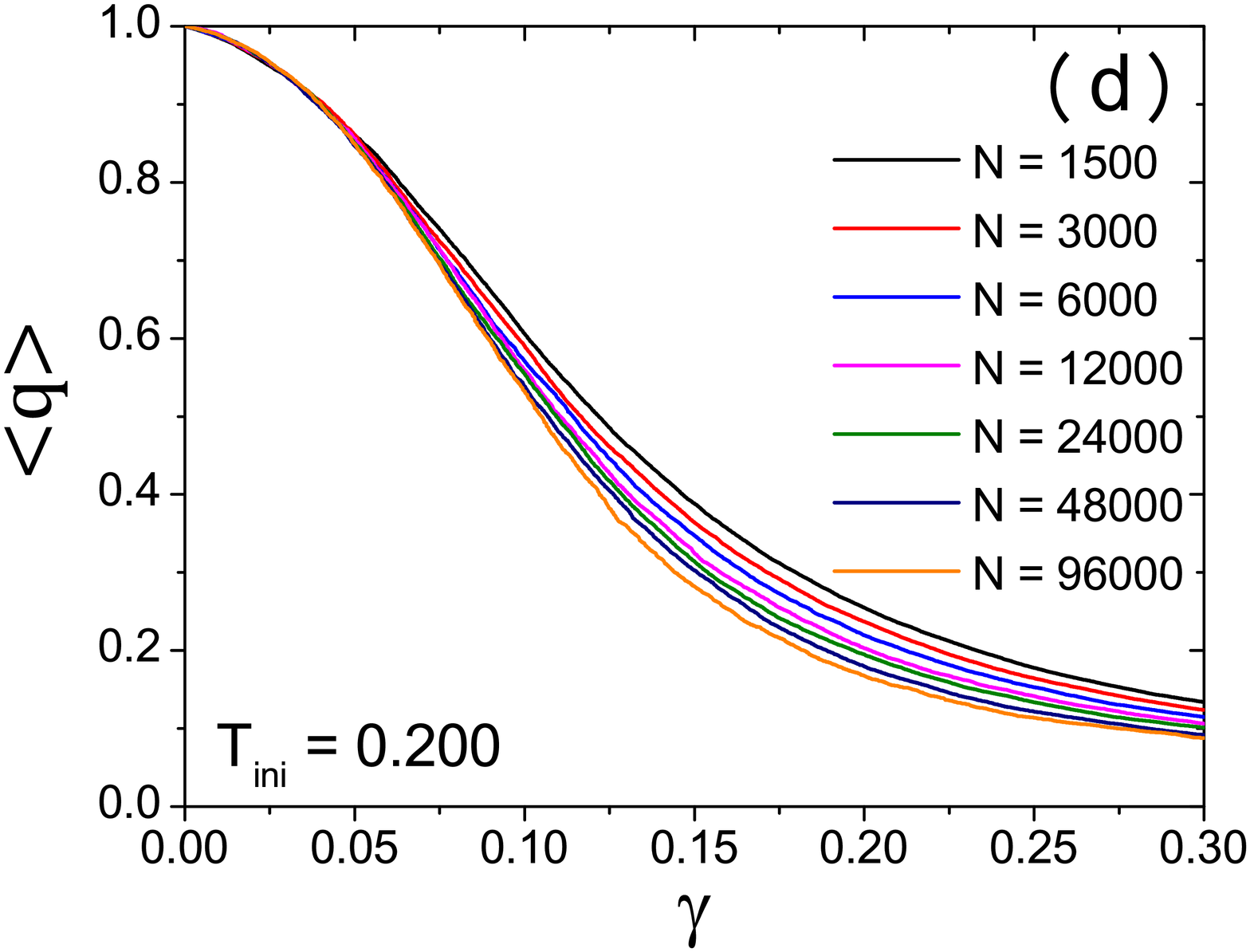}
\includegraphics[width=0.95\columnwidth]{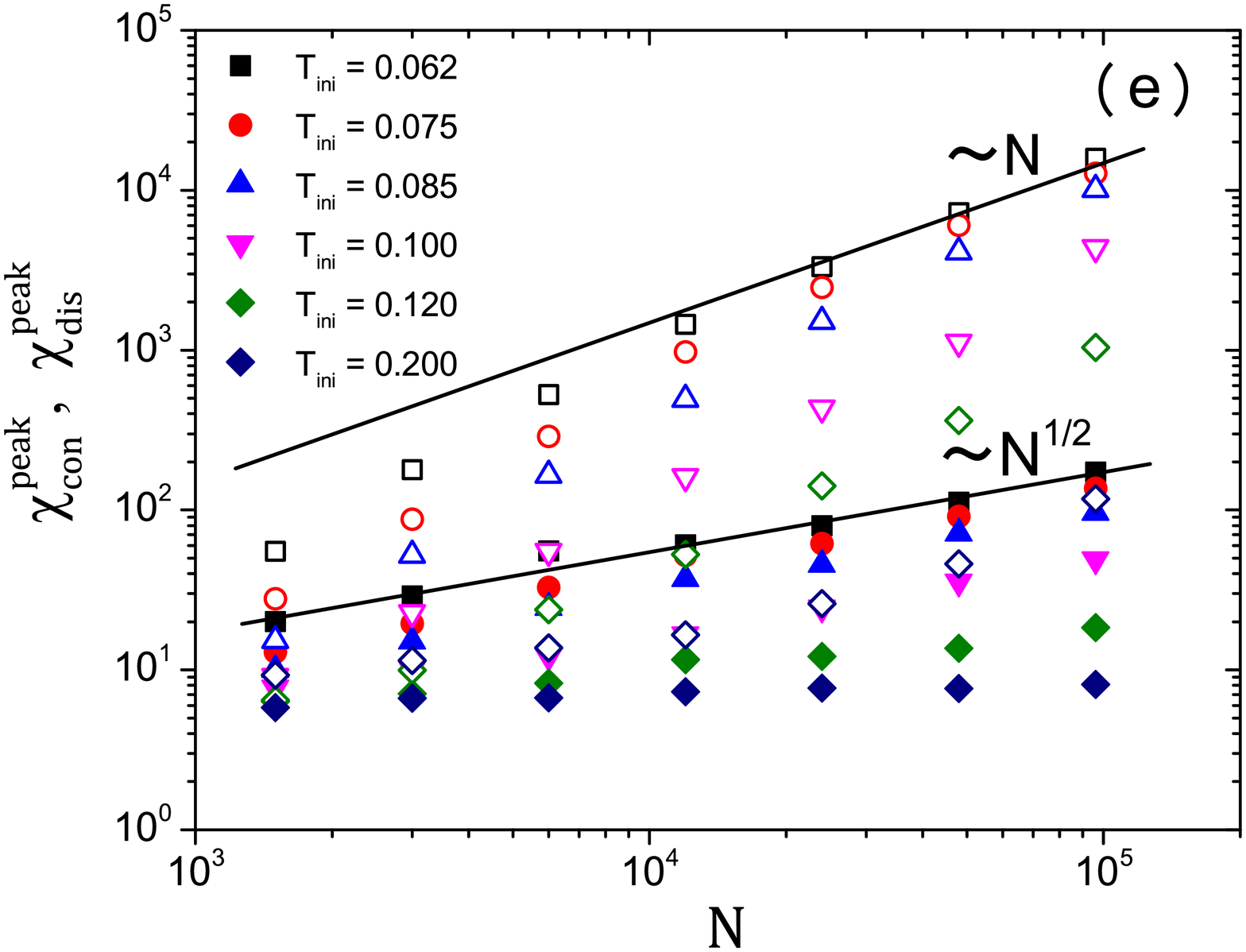}
\caption{
(a-d): Finite size effect of the stress (a,c) and overlap (b,d) for higher $T_{\rm ini}$'s.
(e): $\chi_{\rm con}^{\rm peak}$ (filled symbols) and $\chi_{\rm dis}^{\rm peak}$ (open symbols) obtained from the overlap as a function of $N$. The straight lines correspond to the scaling behavior with $\sim N$ and $\sim N^{1/2}$, respectively.}
\label{fig:highT}
\end{center}
\end{figure}

It has been recently argued that the finite-size effect of $\langle q \rangle$ is always a signature of a discontinuous yielding transition~\cite{procaccia2017mechanical}. At variance with this claim, we show here that a different finite-size behavior is observed at higher $T_{\rm ini}$ where yielding is simply a smooth crossover. We display the system-size dependence of $\langle \sigma \rangle$ and $\langle q \rangle$ for $T_{\rm ini}=0.120$ and $0.200$ in Figs.~\ref{fig:highT}(a-d). Clearly, $\langle q \rangle$ has a stronger dependence on system size than $\langle \sigma \rangle$ for the sizes studied. A possible explanation for this fact is that the relative change of $\langle q \rangle$ during yielding (from $1$ to nearly $0$) is much larger than that of $\langle \sigma \rangle$ (from the maximum stress to the steady state value). Nonetheless, as shown by the behavior of the maximum of the susceptibilities $\chi_{\rm con}$ and $\chi_{\rm dis}$ in Fig.~\ref{fig:highT}(e), $\chi_{\rm con}^{\rm peak}$ and $\chi_{\rm dis}^{\rm peak}$ are suppressed significantly for high $T_{\rm ini}$. Furthermore, the growth of $\chi_{\rm con}^{\rm peak}$ with $N$ tends to saturate at large system size, thereby indicating that yielding is a mere crossover above the critical point. The same behavior is therefore found with the stress and the overlap order parameters.

\section{Avalanche statistics}

\noindent {\it Determination of the stress drops}

The athermal quasi-static shear simulation consists of discrete steps, producing a sequence of values of the stress, $\sigma_i$ ($i=1, 2, \cdots$). To obtain a precise determination of the stress drops, the back-tracking method that proceeds by reducing $\Delta \gamma$ around each stress drop is often used~\cite{karmakar2010statistical}. However, this method is computationally demanding because of the iterative back-trackings. In this paper we use an alternative way to determine the stress drops precisely with fixed $\Delta \gamma$~\cite{zhang2017scaling}. We define the $i$-th stress drop by
\begin{equation}
\Delta \sigma_i =  \sigma_i - (\sigma_{i-1}+ \mu \Delta \gamma),
\end{equation}
where $\mu$ is the locally determined shear modulus~\cite{zhang2017scaling}.
In the above equation, $\sigma_{i-1}+ \mu \Delta \gamma$ is what the stress would be after the strain increment $\Delta \gamma$ if there were no stress drop.

We define a threshold for the stress drops as $\Delta \sigma = -c/N$~\cite{bailey2007avalanche}. We use $c=0.1$. We have checked that the scaling behavior does not change in a range between $c=0.03$ and $3$.
\\

\noindent{\it Avalanche-size distribution and mean size}

\begin{figure}[t]
\begin{center}
\includegraphics[width=0.48\columnwidth]{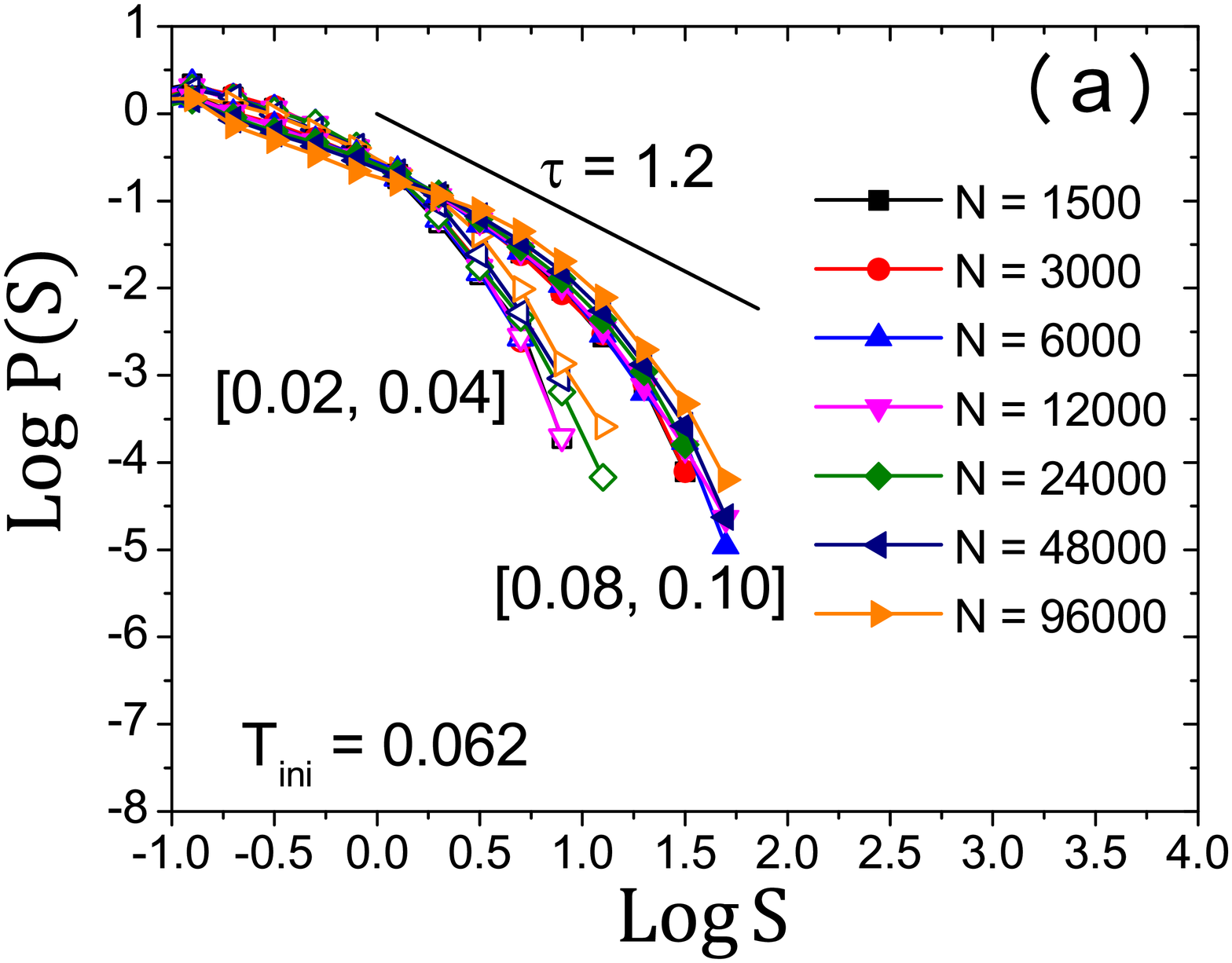}
\includegraphics[width=0.48\columnwidth]{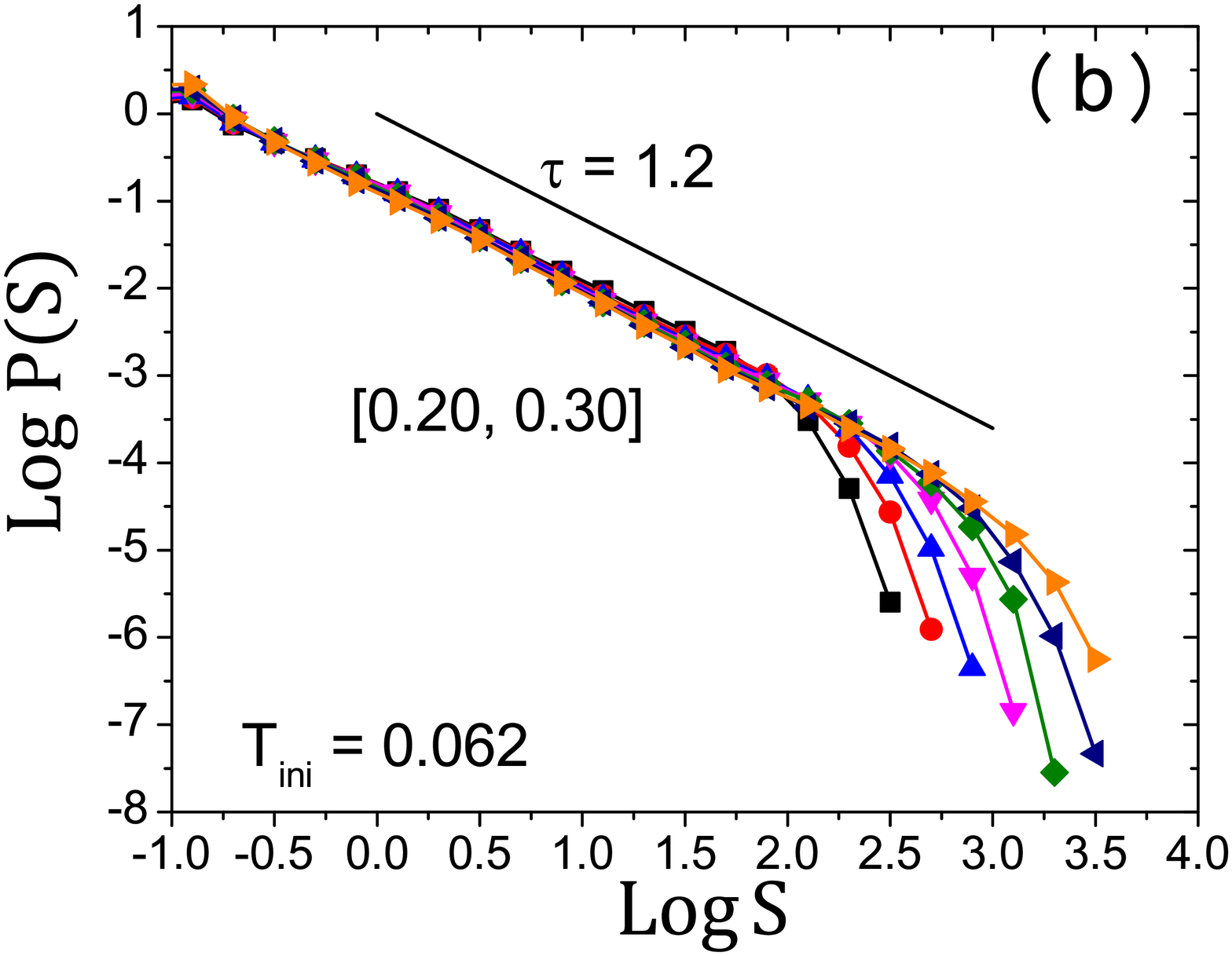}
\includegraphics[width=0.48\columnwidth]{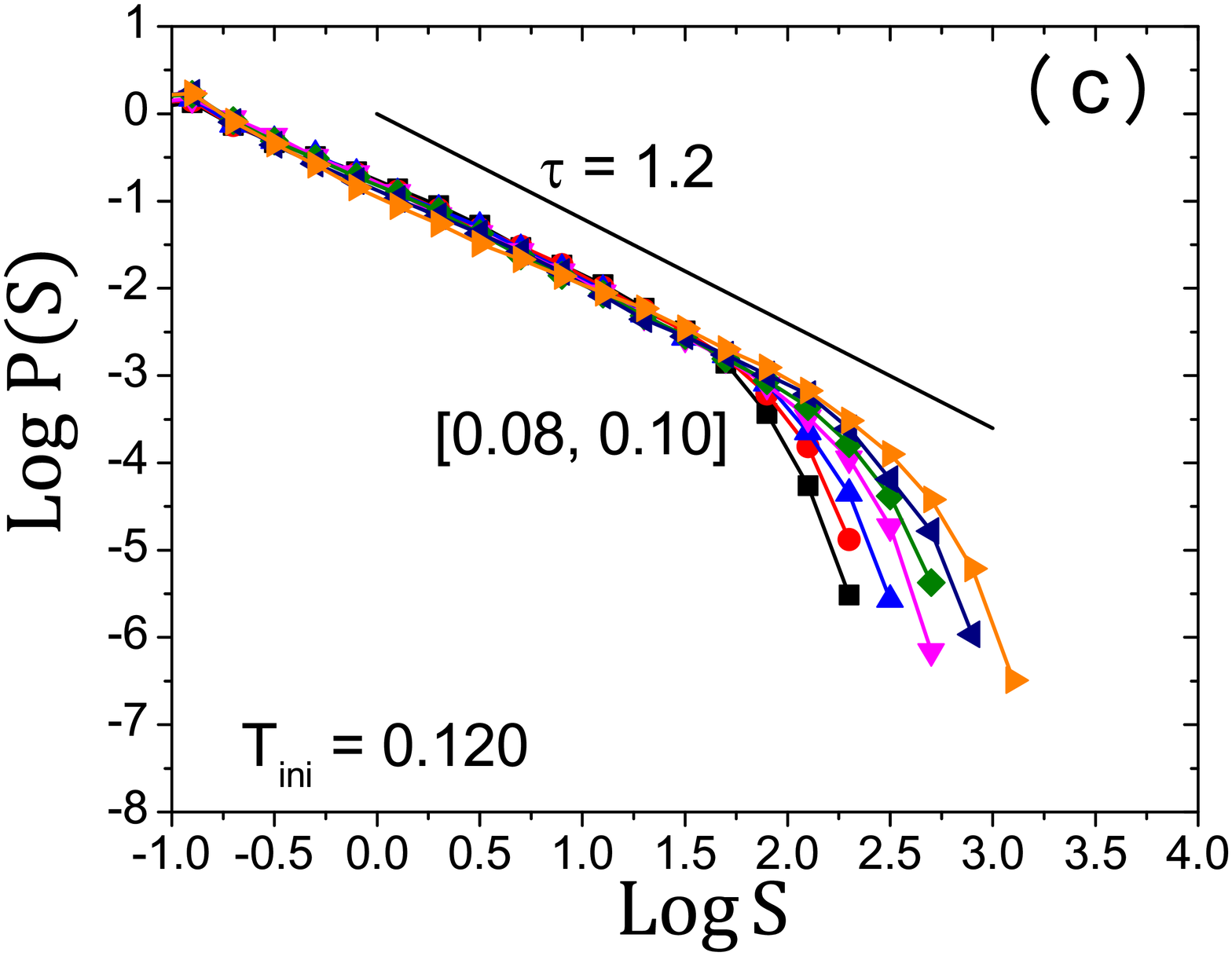}
\includegraphics[width=0.48\columnwidth]{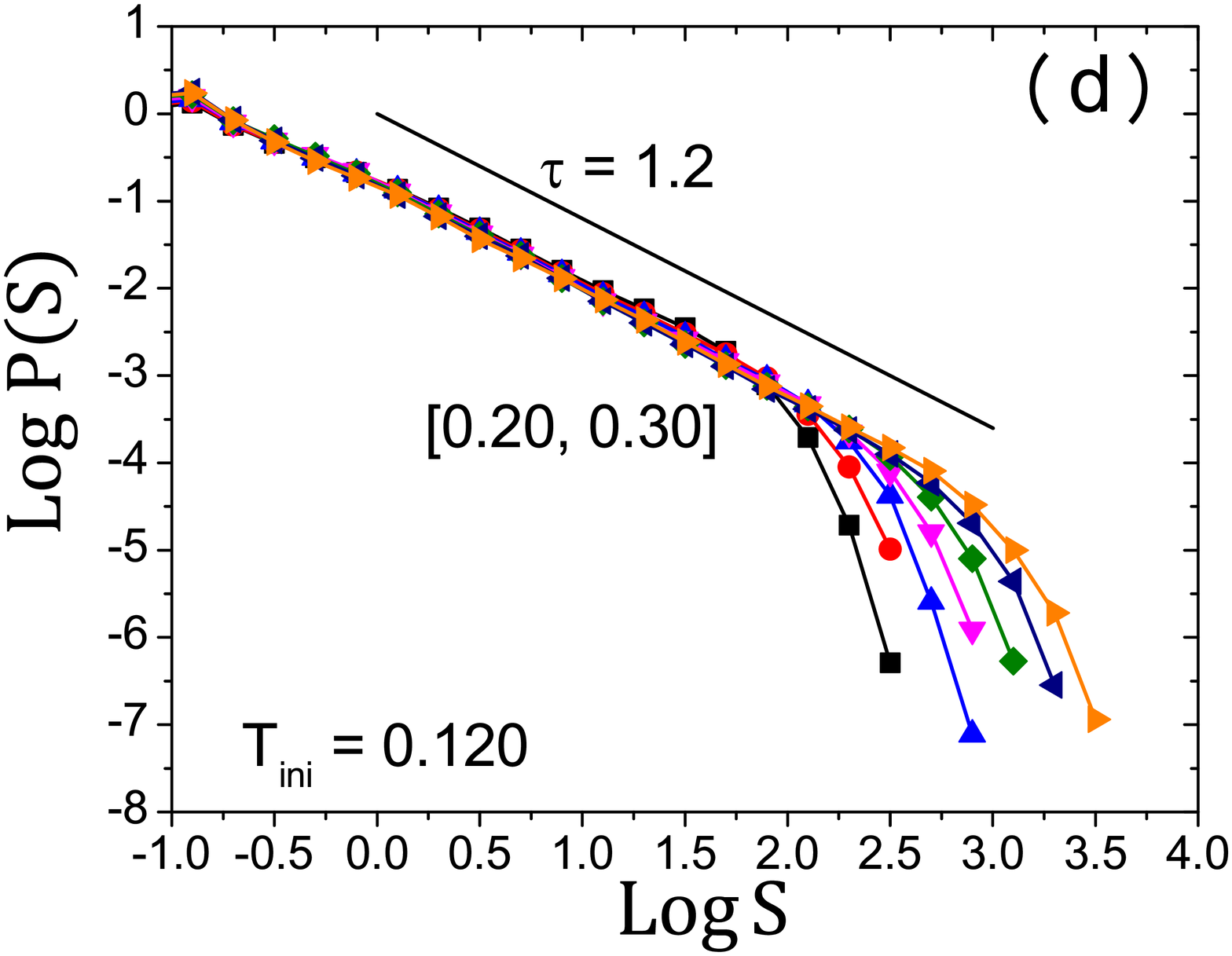}
\caption{
Distribution of the avalanche size $S$.
Top: Before (a) and after (b) yielding for $T_{\rm ini}=0.062$.
In (a) two strain intervals, $\gamma \in [0.02, 0.04]$ (empty points) and $\gamma \in [0.08, 0.10]$ (filled points) are shown.
Bottom: Before (c) and after (d) yielding for $T_{\rm ini}=0.120$.
The straight line corresponds to the power-law decay with $\tau=1.2$.
}
\label{fig:distribution}
\end{center}
\end{figure}

We define the avalanche size $S$ in terms of the stress drop through $S=N|\Delta \sigma|$~\cite{karimi2017inertia}. We measure the distribution $\mathcal P(S)$ for a given interval of $\gamma$ to see the effect of yielding on the avalanche behavior~\cite{karmakar2010statistical,hentschel2015stochastic}. In Fig.~\ref{fig:distribution} we show $\mathcal P(S)$ for temperatures above ($T_{\rm ini}=0.120$) and below ($T_{\rm ini}=0.062$) the critical point, and in each case for deformations  before and after yielding.
After yielding ($\gamma \in [0.20, 0.30]$) for both $T_{\rm ini}=0.062$ and $0.120$ (see Figs.~\ref{fig:distribution}(b) and (d)), $\mathcal P(S)$ behaves as a power-law with a system-size dependent cutoff $S_{\rm cut}$,
\begin{eqnarray}
\mathcal P(S) &\sim& S^{-\tau} f(S/S_{\rm cut}), \label{eq:PS} \\
S_{\rm cut} &\sim& N^{d_f/d},
\label{eq:Scut}
\end{eqnarray}
where $f(x)$ is a monotonically decreasing function and $d_f$ is a fractal dimension. We find that a scaling collapse of the data can be obtained with Eqs.~(\ref{eq:PS}) and (\ref{eq:Scut}) with $\tau \approx 1.2$ and $d_f \approx 1.5$ for this steady-state regime, where the effect of the initial condition has disappeared. These values of $\tau$ and $d_f$ are compatible with other studies~\cite{liu2016}.

Interestingly, a qualitatively different behavior is observed for $T_{\rm ini}=0.062$ and $0.120$ before yielding (see Figs.~\ref{fig:distribution}(a) and (c)). For $T_{\rm ini}=0.062$ a significant suppression of the mean value and of the finite-size dependence is found. In Fig.~\ref{fig:distribution}(a) we show two strain intervals to see the evolution of $\mathcal{P}(S)$ when approaching yielding. $\mathcal{P}(S)$ extends to higher $S$ values while keeping the power-law exponent $\tau \approx 1.2$. 

\begin{figure}[t]
\begin{center}
\includegraphics[width=0.95\columnwidth]{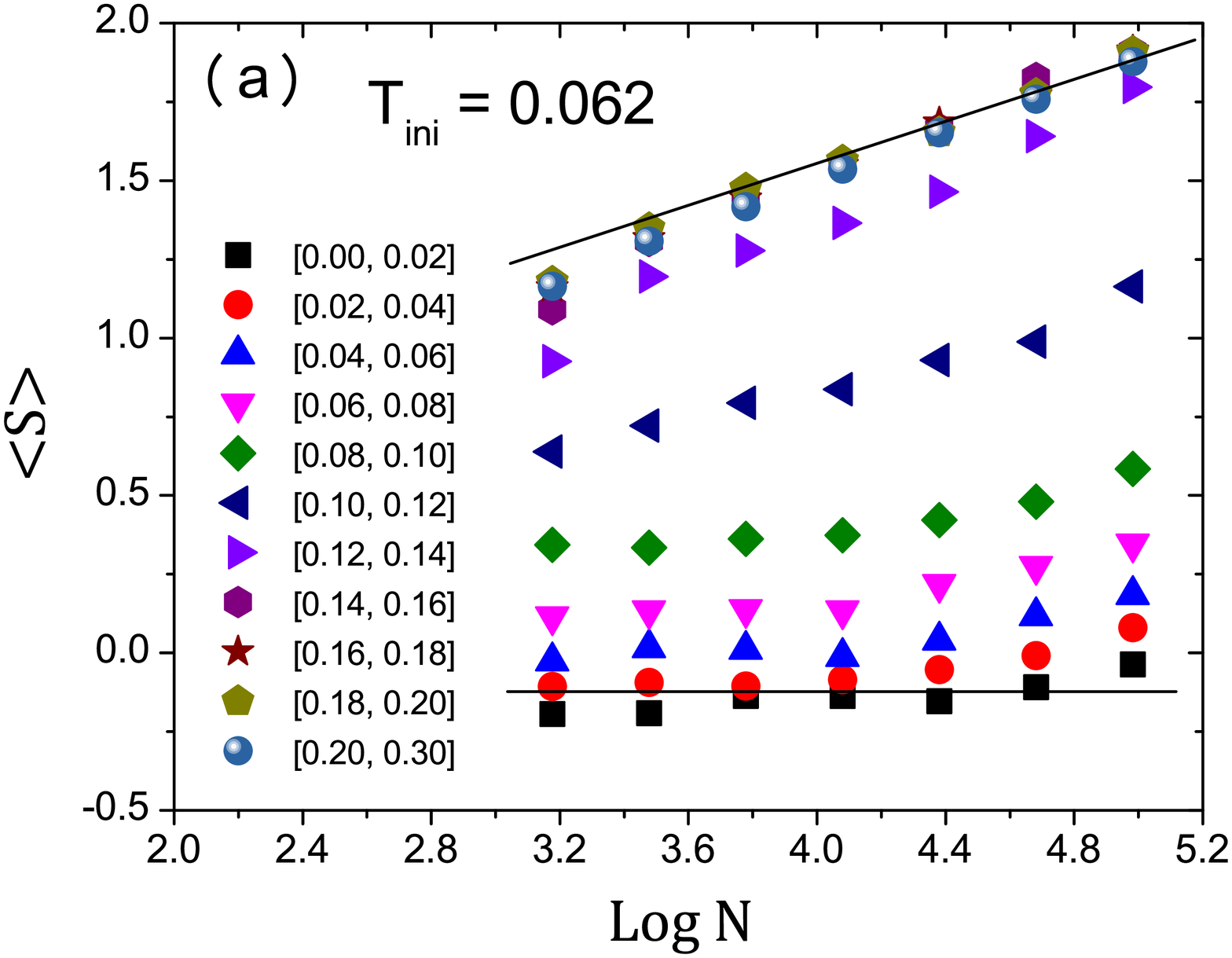}
\includegraphics[width=0.95\columnwidth]{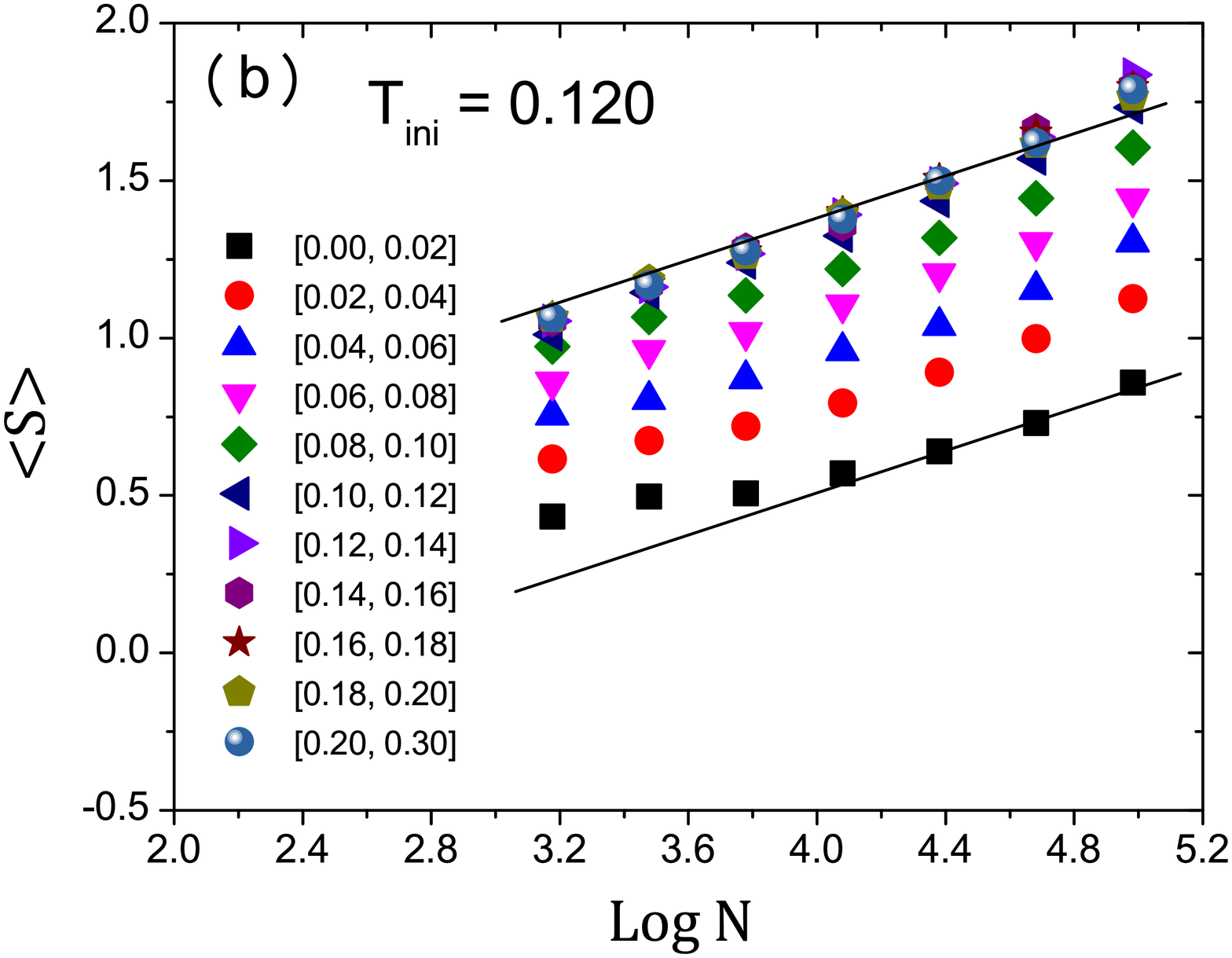}
\caption{
Mean avalanche size versus $N$ as a function of the interval of $\gamma$ for $T_{\rm ini}=0.062$ (a) and $0.120$ (b).
(a): The upper and lower straight lines correspond to $\langle S \rangle \sim N^{1/3}$ and $N^0$, respectively. (b): Both straight lines correspond to $\langle S \rangle \sim N^{1/3}$.
}
\label{fig:mean_S}
\end{center}
\end{figure}

The mean avalanche size $\langle S \rangle$ contains essential information to determine the pseudo-gap exponent $\theta$. According to Refs.~\cite{wyart-rosso,lin2015criticality}, the system-size dependence of $\langle S \rangle$ scales as 
\begin{equation}
\label{eq_first-momentS}
\langle S \rangle \sim N^{\frac{\theta}{1+\theta}},
\end{equation}
at least away from a critical point where the connected susceptibility diverges with system size~\cite{wyart18}. (This additional divergence, however, has no consequence for the way we numerically determine the exponent $\theta$.) This scaling behavior has been confirmed both in steady-state conditions~\cite{wyart-rosso} and before yielding~\cite{lin2015criticality}.
We show $\langle S \rangle$ versus $N$ as a function of the  interval of $\gamma$ for $T_{\rm ini}=0.062$ and $0.120$ in Fig.~\ref{fig:mean_S}. To compute $\langle S \rangle$ we remove the largest stress drop, $\Delta \sigma_{\rm max}$. We can see how $\langle S \rangle$ approaches  the known asymptotic  behavior, $\langle S \rangle \sim N^{1/3}$~\cite{karmakar2010statistical,hentschel2015stochastic}.
Whereas the data for $T_{\rm ini}=0.120$ essentially follow the same scaling behavior $N^{1/3}$ from the beginning of shearing, those for $T_{\rm ini}=0.062$ appear roughly constant at small deformation up to very large system sizes.
 
\section{Marginality of the sheared glass}

\label{marginality}

As shown in the main text, for well annealed glasses, a discontinuous yielding transition separates two distinct regimes. The behavior of the individual samples illustrated in Fig.~2(a) (main text) reveals that the statistics of stress drops is qualitatively different before and after yielding, with fewer and smaller stress drops taking place before yielding. Larger (yet not macroscopic) stress drops seem to appear before yielding as $T_{\rm ini}$ increases. Lin {\it et al.} have recently predicted that an amorphous material quasi-statically sheared at zero temperature in a strain-controlled protocol is marginal at all values of the strain~\cite{lin2015criticality}. The physical reason is the presence of a pseudo-gap in the density of elementary excitations, which means that the distribution $P_{\gamma}(x)$ introduced above behaves as $P_{\gamma}(x) \sim x^\theta$ at $x \to 0$, where $\theta >0$ is an exponent that depends on the strain and reaches a well studied steady-state value ($\theta \approx 0.5-0.6$ in $3d$~\cite{karmakar2010statistical,liu2016}). This marginality implies a scale-free distribution of avalanche sizes and, in particular, a scaling of the average stress drop  as~\cite{wyart-rosso,muller2015marginal,lin2015criticality}  (see also the comment below Eq.~(\ref{eq_first-momentS}))
\begin{equation}
N \langle \Delta \sigma \rangle \sim N^{\theta/(1+\theta)}.
\label{stressdrop}
\end{equation}

\begin{figure}[t]
\centering
\includegraphics[width=0.95\linewidth]{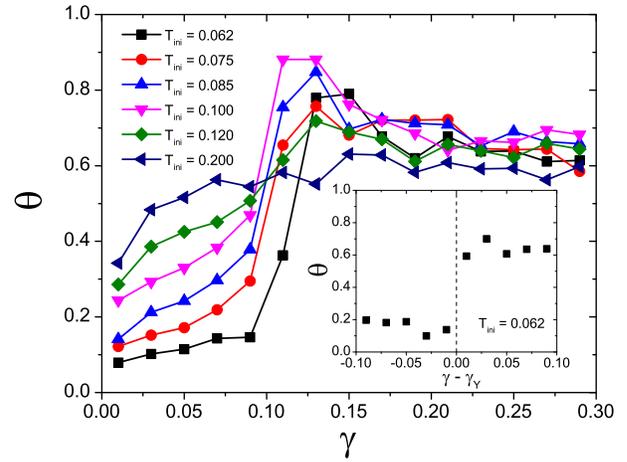}
\caption{Critical exponent $\theta$ versus strain $\gamma$ for various degrees of annealing. The inset show $\theta$ measured as a function of the fluctuating distance to yielding $\gamma_{Y}$ for $T_{\rm ini}=0.062$, which displays a clear discontinuous jump at yielding.}
\label{fig:marginality}
\end{figure}

We have performed a careful analysis of the stress-drop statistics and used Eq.~(\ref{stressdrop}) to extract the exponent $\theta$ as a function of $\gamma$ and $T_{\rm ini}$. As expected from the above qualitative observations, we observe that $\theta$ takes different values before and after yielding (it is smaller before yielding) and that it depends strongly on the preparation temperature $T_{\rm ini}$ before yielding (it is smaller for more stable glasses)~\cite{karmakar2010statistical,lin2015criticality,hentschel2015stochastic}: see Fig.~\ref{fig:marginality}. We always find $\theta>0$, which implies that all the states that we can access display system-spanning plastic events leading to a nontrivial size dependence of the  stress drops. However, for the most stable glasses, $\theta$ becomes very small at small deformation, $\theta \sim 0.1$, indicating that criticality is very weak in these samples, which are thus very close to being perfect elastic solids~\cite{hentschel2015stochastic}. 

These findings fit well in our analysis, since we expect the discontinuous yielding transition to be associated with a discontinuous variation of $\theta$~\cite{hentschel2015stochastic,leishangthem2017yielding} across yielding. Our data for the lowest $T_{\rm ini}$ are consistent with this expectation. The discontinuous jump of $\theta$ can be evinced if $\theta$ is measured together with $\gamma-\gamma_{Y}$, where $\gamma_{Y}$ is the location of $\Delta \sigma_{\rm max}$ determined for each sample. Determining $\theta$ as a function of the fluctuating distance to yielding, as shown in the inset of Fig.~\ref{fig:marginality}, provides strong support for a discontinuity in $\theta$ at low $T_{\rm ini}$. By contrast, $\theta$ evolves smoothly with $\gamma$ toward its steady-state value for large $T_{\rm ini}$. Interestingly, between these two extreme situations, for the two intermediate temperatures close to the critical point, $T_{\rm ini}=0.085$ and $T_{\rm ini}=0.100$, the value of $\theta$ changes rapidly in the region of the yielding transition where it passes through a large maximum. The latter arises due to the large fluctuations of the stress drops, likely associated with the criticality analyzed in the main text. This behavior provides additional evidence for the presence of an annealing-controlled random critical point.

\section{Importance of rare fluctuations}

An important consequence of the scenario proposed to account for the brittle yielding phase is that it is similar to the physics of spinodal instabilities in the presence of quenched disorder and finite dimensions, a situation recently explored in the context of the RFIM~\cite{nandi}. A key aspect is the crucial role played by rare regions of the sample that may trigger the instability. As a result, very large finite-size effects are expected. It is possible to investigate the influence of these rare fluctuations numerically, by inserting in the sample "defects" or "seeds" that would otherwise be rare if spontaneously nucleated. 
For the present paper we simply report the following preliminary numerical experiment. 

\begin{figure}[t]
\begin{center}
\includegraphics[width=0.4\columnwidth]{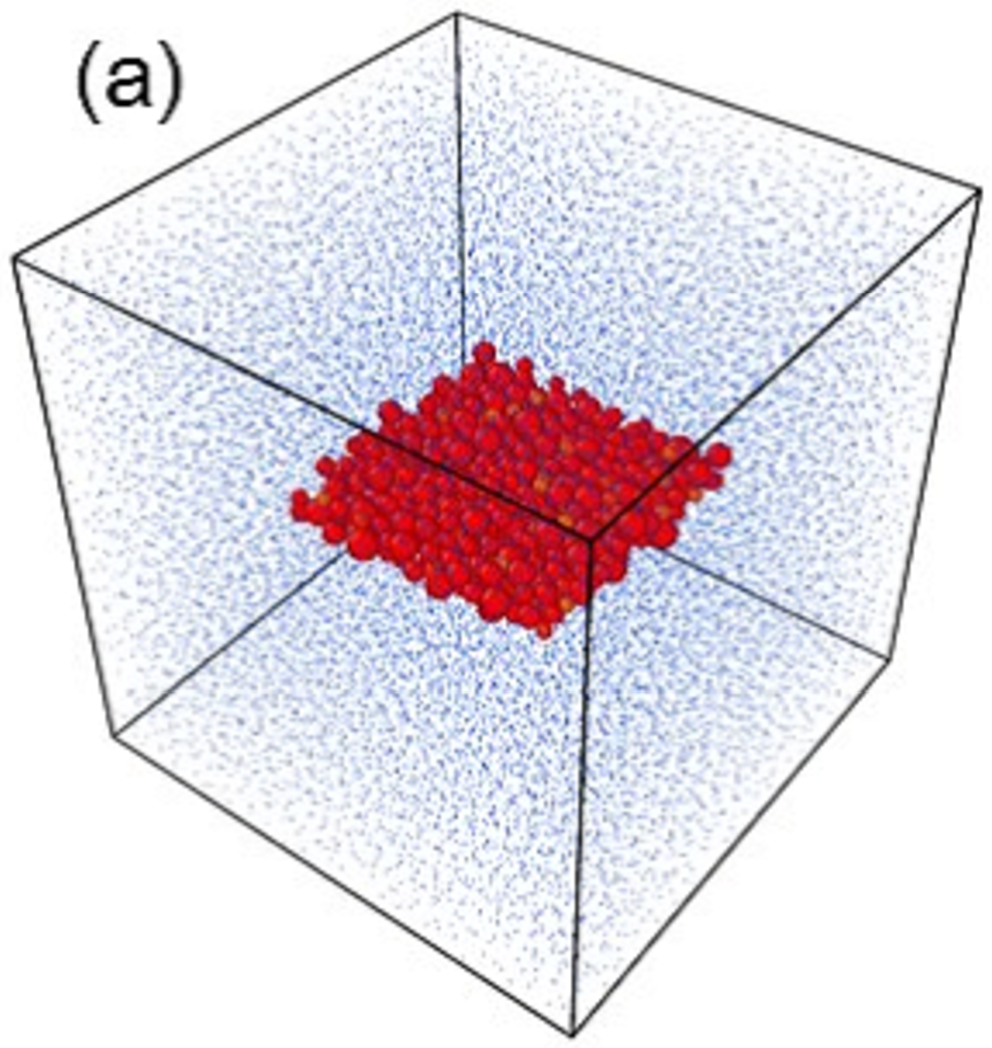}
\includegraphics[width=0.55\columnwidth]{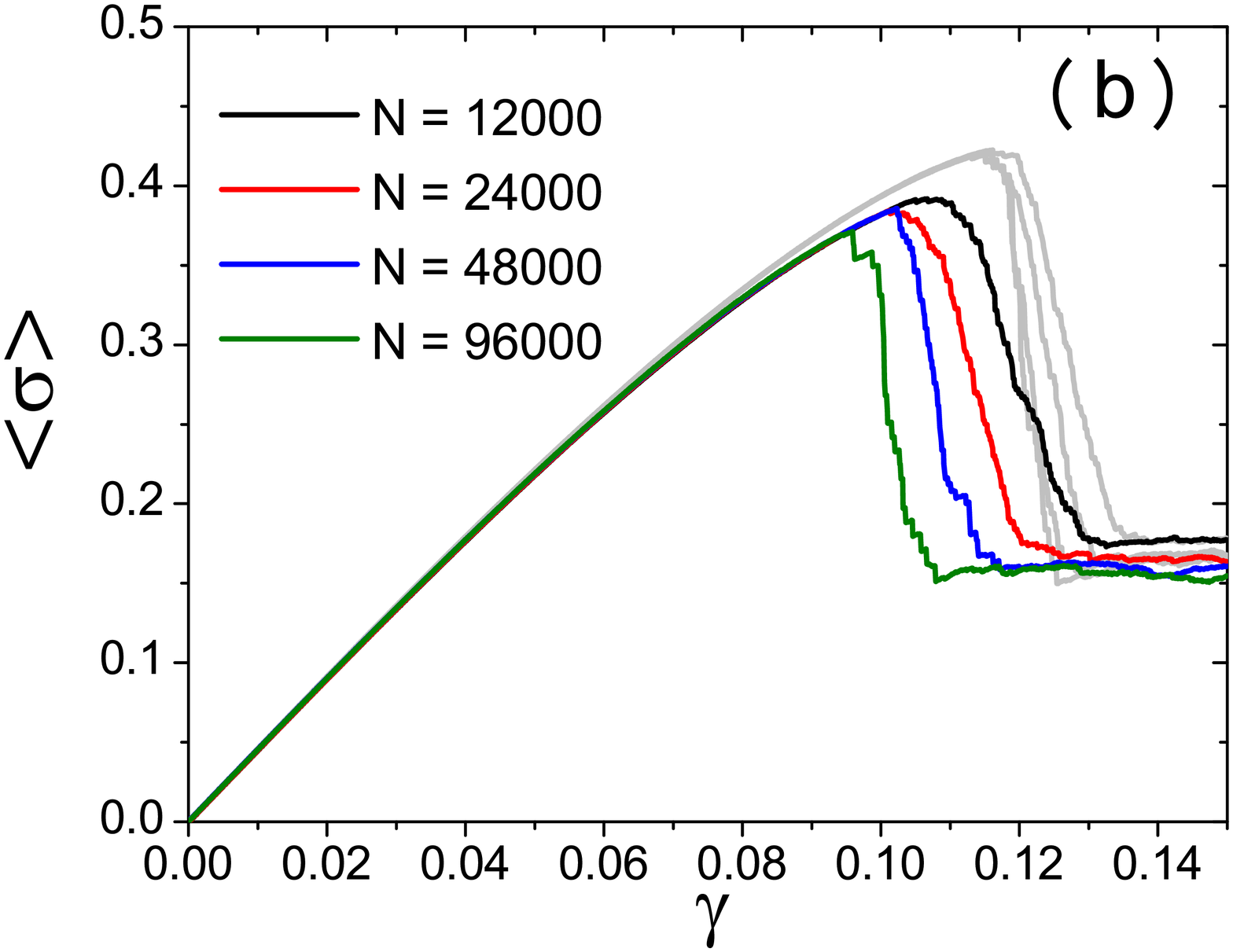}
\caption{(a) A snapshot ($N=24000$) describing the idea of inserting a defect  (here an essentially planar seed of size $L/2 \times L/2 \times 2$) of poorly annealed material (red particles) inside a well annealed sample (blue particles with decreased diameter). The defect is inserted with its short side perpendicular to the shear direction. (b) Shift of the yielding transition for a defect of $L/2 \times L/2 \times 2$ inserted in glass samples  of increasing system size at $T_{\rm ini}=0.062$. The original (without defect) stress-strain curves for $N=12000-96000$ are shown in grey.}
\label{fig:seed}
\end{center}
\end{figure}

To create a defect (soft region) of a given geometry and size inside a well-annealed glass sample prepared at a temperature $T_{\rm ini}=0.062$, we reheat by Monte Carlo simulation the particles belonging to the defect at a high temperature $T=0.3$, leaving fixed the position of all other particles, as illustrated in Fig.~\ref{fig:seed}(a). After this step, the system is quenched down to zero temperature again by the conjugate-gradient method.  
We then use these glass samples with a defect as initial condition for the deformation protocol conducted as before. We typically find that the presence of such a defect shifts the location of the yielding transition. This is illustrated in Fig.~\ref{fig:seed}(b), where we show the stress-strain curve for a seed of $L/2 \times L/2 \times 2$ flat plate inside a sample of volume $V=L^3$. Our data suggest that as one increases the system size, the location of the yielding transition changes, in qualitative agreement with the analysis in Ref.~\cite{nandi}. More work along these lines is in progress.        

\section{Movies}

\begin{figure}[t]
\begin{center}
\includegraphics[width=0.95\columnwidth]{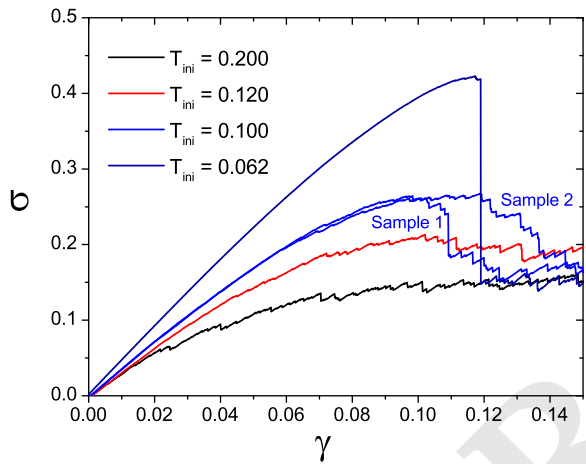}
\caption{Stress-strain curves for the samples shown in the movies. For $T_{\rm ini}=0.100$ we show two samples with relatively brittle (Sample 1) and ductile (Sample 2) behaviors.}
\label{fig:movie}
\end{center}
\end{figure}

Five movies of some representative samples are available online. The sample size is $N=96000$. The color bar corresponds to the non-affine displacement from the origin, $|{\bf r}_i^{\rm NA} (\gamma)-{\bf r}_i(0)|$, and we show four preparation temperatures. The corresponding stress-strain curves for all movies are shown in Fig.~\ref{fig:movie}.

1) \verb|Tini0062.mp4|
The preparation temperature is $T_{\rm ini}=0.062$, and the system has a sharp discontinous yielding transition.

2) \verb|Tini0100_sample1.mp4|
The preparation temperature is $T_{\rm ini}=0.100$, close to the critical point, and the system has a sharp discontinous yielding transition.

3) \verb|Tini0100_sample2.mp4|
The preparation temperature is $T_{\rm ini}=0.100$, close to the critical point and the system has a smooth ductile yielding transition.

4) \verb|Tini0120.mp4|
The preparation temperature is $T_{\rm ini}=0.120$, and the system has a smooth stress overshoot. 

5) \verb|Tini0200.mp4|
The preparation temperature is $T_{\rm ini}=0.200$, and the system has a monotonic stress strain curve. 

\bibliography{pnas-sample.bib}

\end{document}